\documentclass[aps,prd,twocolumn,superscriptaddress,nofootinbib]{revtex4-1}%
\usepackage{epsf,epsfig}
\usepackage[utf8]{inputenc}
\usepackage{amssymb,amsmath,amsfonts}
\usepackage{color}
\usepackage{slashed}
\usepackage{tensor}
\usepackage{braket}
\usepackage[colorlinks=true]{hyperref}  
\hypersetup{
    bookmarks=true,         
    unicode=false,          
    pdftoolbar=true,        
    pdfmenubar=true,        
    pdffitwindow=false,     
    pdfstartview={FitH},    
    colorlinks=true,       
    linkcolor=magenta, 
    citecolor=blue,        
    filecolor=magenta,      
    urlcolor=cyan           
} 

\newcommand{\bea}{\begin{eqnarray}}
\newcommand{\eea}{\end{eqnarray}}
\newcommand{\ba}{\begin{eqnarray}}
\newcommand{\ea}{\end{eqnarray}}

\newcommand{\beq}{\begin{equation}}
\newcommand{\eeq}{\end{equation}}
\newcommand{\beqa}{\begin{eqnarray}}
\newcommand{\eeqa}{\end{eqnarray}}
\newcommand{\beqar}{\begin{eqnarray*}}
	\newcommand{\eeqar}{\end{eqnarray*}}

\newcommand{\R}{\mathcal{R}}

\newcommand{\E}{\mathcal{E}}

\newcommand{\cO}{\mathcal{O}}

\usepackage{color}

\newcommand{\req}[1]{(\ref{#1})} 

\begin{document}
\allowdisplaybreaks

\title{Novel higher-curvature variations of $R^2$ inflation}

\author{Pablo A. Cano}
\email{pabloantonio.cano@kuleuven.be}
\affiliation{Institute for Theoretical Physics, KU Leuven. Celestijnenlaan 200D, B-3001 Leuven, Belgium}

\author{Kwinten Fransen}
\email{kwinten.fransen@kuleuven.be}
\affiliation{Institute for Theoretical Physics, KU Leuven. Celestijnenlaan 200D, B-3001 Leuven, Belgium}

\author{Thomas Hertog}
\email{thomas.hertog@kuleuven.be}
\affiliation{Institute for Theoretical Physics, KU Leuven. Celestijnenlaan 200D, B-3001 Leuven, Belgium}


\begin{abstract}
We put forward novel extensions of Starobinsky inflation, involving a class of `geometric' higher-curvature corrections that yield second-order Friedmann-Lema\^itre equations and second-order-in-time linearized equations around cosmological backgrounds. We determine the range of models within this class that admit an extended phase of slow roll inflation as an attractor. By embedding these theories in anti-de Sitter space, we derive holographic `unitarity' bounds on the two dominant higher-order curvature corrections. Finally we compute the leading corrections to the spectral properties of scalar and tensor primordial perturbations, including the modified consistency relation $r=-8n_{T}$. Remarkably, the range of models singled out by holography nearly coincides with the current observational bounds on the scalar spectral tilt. Our results indicate that future observations have the potential to discriminate between different higher-curvature corrections considered here.

\end{abstract}

\maketitle

\section{Introduction}

Starobinsky realised long ago that the trace anomaly of a large number of light matter fields can support a de Sitter phase in the early universe \cite{Starobinsky:1980te,Starobinsky:1983zz}. The de Sitter solution is unstable but can be long-lived, and decays via slow roll into a matter dominated Friedman-Lema\^itre Robertson-Walker (FLRW) universe. Starobinsky's motivation was to demonstrate that quantum effects of matter fields might resolve the big bang singularity. From a modern perspective it is interesting that the conformal anomaly might have been the source of a finite but significant period of slow roll inflation in the early universe. 

Starobinsky showed that the de Sitter solution supported by the trace anomaly is unstable both to the future and to the past, so it was not clear how the universe could have entered the de Sitter state. However, this problem can be overcome by an appeal to quantum cosmology \cite{Hartle:1983ai,Vilenkin:1985md}. Specifically, the semiclassical no-boundary wave function predicts that a classical expanding universe with a large number of conformal matter fields emerges through the de Sitter phase Starobinsky identified \cite{Hawking:2000bb,Hartle:2007gi,Hartle:2008ng}.

Remarkably, the predictions of Starobinsky inflation for the fluctuations in the cosmic microwave background (CMB) are in excellent agreement with current observations \cite{Akrami:2018odb}. If the present horizon scale left during the slow roll phase then the phenomenology is to a large extent determined by the $R^2$ term in the action, which acts as an inflaton and is responsible for the instability of the de Sitter solution in the first place. For this reason, the original Starobinsky model is often reduced to $R^2$ inflation. Nevertheless, it has long been known that other higher derivative terms featuring in the trace anomaly as well as the non-local effects it gives rise to, affect the details of the pattern of primordial perturbations (see e.g. \cite{Starobinsky:1980te,Starobinsky:1983zz,Hawking:2000bb}). Moreover since this class of models is most naturally viewed in the context of an effective field theory (EFT) expansion around General Relativity (GR), there is no justification to exclude on ad hoc grounds yet further higher derivative contributions to the action.  

These considerations are no longer of merely theoretical interest since the next generation of CMB experiments has the potential to unlock this region of parameter space, opening up the bright prospect to observationally differentiate between variations of trace anomaly inflation. Thus there is a strong motivation to further explore this class of inflationary models, as theoretically appealing and observationally viable alternatives to scalar field driven inflation. What is the space of higher-derivative theories in the neighbourhood of the Starobinsky model that admit inflation, and what is their phenomenology?

In this spirit, higher-derivative theories with $R^3$ and $R\nabla^2 R$ terms have been considered \cite{Gottlober:1989ww,Berkin:1990nu,Amendola:1993bg,Cuzinatto:2013pva,Castellanos:2018dub}
as well as more general modifications of the form $f(R)$ and  $f(R,\nabla^2 R)$ (See \textit{e.g.} resp. \cite{Huang:2013hsb,Sebastiani:2013eqa,Bamba:2015uma} and  \cite{Carloni:2018yoz}). These are somewhat special, however, since models of this kind are equivalent to scalar-tensor theories with second-order equations of motion \cite{Sotiriou}. On general grounds one expects an EFT expansion of GR to include also a tower of operators containing explicit Ricci and Riemann tensors. In this paper we study the viability and phenomenological implications of adding a particular class of such more general terms as perturbative extensions to $R^2$ inflation.

The specific higher-curvature Lagrangians we consider were identified in \footnote{The non-local infinite-derivative gravities of \cite{Biswas:2011ar} are another family of theories intended to avoid ghosts and instabilities. These were considered as modifications of $R+R^2$ inflation in \cite{Craps:2014wga,Koshelev:2016xqb}.} \cite{Arciniega:2018fxj,Cisterna:2018tgx,Arciniega:2018tnn}. A distinctive property of this family of theories is that they give rise to second-order Friedmann-Lema\^itre equations and, at least some, possess linearized equations that are second order in time derivatives \cite{Cisterna:2018tgx}. These properties render possible a comprehensive  analysis of their cosmological backgrounds and of the evolution of perturbations. In the context of $R^2$ inflation the effect of these higher-curvature terms is, essentially, to modify the response of the spacetime geometry to the matter content. For this reason we call these `geometric' corrections, as opposed to the $R^2$ term.\footnote{In \cite{Arciniega:2018tnn} it was shown that higher-curvature Lagrangians of the form \cite{Arciniega:2018fxj,Cisterna:2018tgx,Arciniega:2018tnn} can lead to inflation in a radiation-dominated universe without an $R^2$ term or an inflaton field --- a scenario that was aptly called \emph{Geometric inflation}. However, in \cite{Edelstein:2020nhg} it was shown that a significant period of inflation in this case requires transplanckian radiation or matter densities rendering this scenario questionable. A possible way out is to include a scalar field  \cite{Edelstein:2020lgv}. See also \cite{Erices:2019mkd,Arciniega:2020pcy,Marciu:2020ysf,Quiros:2020uhr,Marciu:2020ski,Quiros:2020eim} for other follow-ups of the geometric inflation scenario.}

We determine the range of models within this class that admit an extended phase of slow roll inflation as an attractor solution. We also evaluate the semiclassical no-boundary wave function of the universe in the presence of the higher-curvature corrections and show that its prediction of an inflationary origin is robust. By embedding these theories in anti-de Sitter space, we also derive holographic `unitarity' bounds on the two dominant higher-order curvature corrections and argue these translate into constraints on the theories in the de Sitter context too. We find that these constraints require that the couplings of the higher-order terms must be much smaller than the coefficient of the $R^2$ term, providing a theoretical ground for the scale separation between these couplings that we adopt. 
Nevertheless, $\mathcal{O}(1)$ effects are still allowed by these bounds at the beginning of inflation. Taking this into account, we investigate the effect of the leading cubic and quartic higher-curvature corrections on the spectral properties of the tensor and scalar perturbations in $R^2$ inflation. We compute in particular the changes to the tensor to scalar ratio $r$, the scalar tilt $n_s$ and the consistency relation $r=-8n_{T}$ and quantify the promising prospects to observationally discriminate between these variations of $R^2$ inflation.

\section{Inflationary Cosmology in Higher-curvature gravity}\label{sec:cosmohigher}

The `geometric' higher-curvature corrections to $R^2$ inflation that we consider, contain terms of the form Riem$^n$ and Ricci$^n$. We first introduce these higher-curvature theories an sich and then discuss their role as novel extensions of $R^2$ inflation.

\subsection{Geometric higher-curvature terms}

In recent years, a new class of higher-curvature Lagrangians known as \emph{Generalized Quasi-topological Gravities} (GQTGs) has been identified \cite{PabloPablo,Hennigar:2016gkm,PabloPablo2,Hennigar:2017ego,PabloPablo3,Ahmed:2017jod,Bueno:2019ycr}. These theories are characterized by having second-order linearized equations on maximally symmetric backgrounds and by allowing for single-function generalizations of the Schwarzschild black hole whose thermodynamic properties can be obtained exactly. We refer to \cite{Hennigar:2017ego,PabloPablo3} for more precise information regarding the definition and properties of these theories.  
It is worth pointing out that, when considering perturbative couplings, GQTGs may provide a basis for an EFT expansion of vacuum GR \cite{Bueno:2019ltp}. Thus, by working with GQTGs one does not lose any information on the EFT of gravity and in turn one is able to perform exact computations rather than perturbative or approximate ones. 

With regard to cosmology, Ref. \cite{Arciniega:2018fxj} found that at the cubic level in the curvature a particular GQTG leads to second-order Friedmann-Lema\^itre equations. This result was extended in Ref.~\cite{Arciniega:2018tnn} to higher orders in the curvature and it was argued that Lagrangians of this kind exist at all orders. Besides, it was reported in Ref.~\cite{Cisterna:2018tgx} that the linear perturbations of the cubic theory around FLRW backgrounds satisfy equations of motion of second order in time derivatives. Here we show the existence of theories with analogous properties at higher orders.

These features make this class of theories appealing for cosmological studies. For general higher-order gravities the presence of higher-order time derivatives sources short-scale instabilities and one must necessarily resort to perturbative schemes. By contrast, in these theories the absence of higher-order time derivatives allows one, at least in principle,\footnote{As recently noted in Refs.~\cite{Pookkillath:2020iqq,Jimenez:2020gbw}, these theories may suffer from other types of instabilities associated to strong coupling issues. We comment on this below.} to perform computations without having to restrict to any approximations. In essence, the only effect of these Lagrangians is to modify the response of the geometry to the stress-energy. 

On a different note, let us mention that the Friedmann-Lema\^itre equations of these theories turn out to be remarkably similar to those found in duality-invariant string cosmology with $\alpha'$ corrections \cite{Hohm:2019jgu}. This suggests that this family of theories might have some connection with string theory which would be worth exploring elsewhere.

Thus, at each order in the curvature $n$ we consider higher-derivative densities $\mathcal{R}_{(n)}$ satisfying the following properties:\footnote{The theories with second-order cosmological equations and those of the GQT class in principle form two independent sets with a non-zero overlap. Although for cosmology we only need those of the first set, choosing theories belonging to the intersection brings additional advantages. For instance the holographic dictionary of GQT theories is known --- see section \ref{sec:holoconst}.}
\begin{enumerate}
\item They belong to the Generalized Quasi-topological class
\item They yield second-order Friedmann-Lema\^itre (FL) equations
\item The linearized equations on FLRW backgrounds are of second order in time derivatives
\end{enumerate}

At each order in the curvature, there are usually several non-trivial densities satisfying these requirements and the number of these densities grows with $n$. However, as noted in \cite{Arciniega:2018tnn}, the contribution of these densities to the FL equations is degenerate, in the sense that the contributions from all the densities of a given order $n$ are proportional to one another. On the other hand, the degeneracy will be broken when considering general perturbations. Thus, the terms $\mathcal{R}_{(n)}$ will be composed of the sum of several densities, but we can specify an overall normalization that will determine their contribution to the FL equations. We shall choose a normalization such that
\begin{equation}\label{eq:normR}
\bar{\mathcal{R}}_{(n)}=-\frac{12}{n-2}(-K)^n\, ,
\end{equation}
when evaluated on $\bar{R}_{\mu\nu\rho\sigma}=2K g_{\mu[\rho}g_{\sigma]\nu}$.

The Einstein-Hilbert term $R$ obviously satisfies the requirements 1-3 and it is normalized according to this rule up to a sign. At quadratic order, the Gauss-Bonnet density $\mathcal{X}_{4}$ is the only Lagrangian satisfying the conditions 1-3, but it is topological and hence does not modify the equations of motion. The cubic correction is somewhat canonical and it has often been called ``cosmological Einsteinian cubic gravity'' \cite{Arciniega:2018fxj}. This density is usually expressed as follows
\begin{equation}\label{eq:cECG}
\mathcal{R}_{(3)}=-\frac{1}{8}\left(\mathcal{P}-8\mathcal{C}\right)\, ,
\end{equation}
where $\mathcal{P}$ and $\mathcal{C}$ are given by

\begin{align}\notag
\mathcal{P}=&12\tensor{R}{_{\mu}^{\rho}_{\nu}^{\sigma}}\tensor{R}{_{\rho}^{\delta}_{\sigma}^{\gamma}}\tensor{R}{_{\delta}^{\mu}_{\gamma}^{\nu}}+\tensor{R}{_{\mu\nu}^{\rho\sigma}}\tensor{R}{_{\rho\sigma}^{\alpha\beta}}\tensor{R}{_{\alpha\beta}^{\mu\nu}}\\
&-12R_{\mu\nu\rho\sigma}R^{\mu\rho}R^{\nu\sigma}+8\tensor{R}{_{\mu}^{\nu}}\tensor{R}{_{\nu}^{\rho}}\tensor{R}{_{\rho}^{\mu}}\, ,\\\notag
\mathcal{C}=&R_{\mu\nu\rho\sigma}\tensor{R}{^{\mu\nu\rho}_{\lambda}}R^{\sigma\lambda}-\frac{1}{4}R_{\mu\nu\rho\sigma}R^{\mu\nu\rho\sigma}R-2R_{\mu\nu\rho\sigma}R^{\mu\rho}R^{\nu\sigma}\\
&+\frac{1}{2}R_{\mu\nu}R^{\mu\nu}R\, .
\end{align}
The density $\mathcal{P}$ is the one corresponding to Einsteinian cubic gravity \cite{PabloPablo}, while the term $\mathcal{C}$ was identified in \cite{Hennigar:2017ego}, and both belong to the GQT class. Unlike $\mathcal{P}$, it turns out that $\mathcal{C}$ does not contribute to the equations of motion for static spherically symmetric solutions, but nevertheless it contributes to the FL equations, and it is this precise combination of $\mathcal{P}$ and $\mathcal{C}$ that yields second-order equations. An example of a quartic theory satisfying the points 1 and 2 can be found in \cite{Arciniega:2018tnn} but the condition 3 was not taken into account. Here we have extended that analysis by searching for the most general quartic Lagrangian satisfying all three conditions. We have found that there are three independent Lagrangians $\mathcal{R}_{(4)}^{A}$, $\mathcal{R}_{(4)}^{B}$ and $\mathcal{R}_{(4)}^{C}$. There are infinite forms of expressing these densities because out of the 26 possible quartic Riemann invariants only 13 of them are linearly independent in $D=4$, but eliminating redundant terms \cite{0264-9381-9-5-003} we can write them as follows
\begin{widetext}
\begin{align}
\notag
\mathcal{R}^{A}_{(4)}=&\frac{3}{32}R^{\mu\nu\rho\sigma }\tensor{R}{_{\mu\nu }^{\delta\gamma }}\tensor{R}{_{\delta\gamma }^{\chi\xi}}R_{\rho\sigma \chi\xi}-\frac{1}{16}(R_{\mu\nu\rho\sigma }R^{\mu\nu\rho\sigma })^2+\frac{1}{6}R\tensor{R}{_{\mu}^{\rho}_{\nu}^{\sigma}}\tensor{R}{_{\rho}^{\delta}_{\sigma}^{\gamma}}\tensor{R}{_{\delta}^{\mu}_{\gamma}^{\nu}}
-\frac{3}{4}R^{\mu\nu }R^{\rho\sigma }\tensor{R}{^{\delta\gamma }_{\mu\rho}}R_{\delta\gamma \nu \sigma}\\\notag
&+\frac{1}{16}R^2R_{\mu\nu\rho\sigma }R^{\mu\nu\rho\sigma}+\frac{1}{2}RR_{\mu\nu\rho\sigma }R^{\mu\rho}R^{\nu\sigma}-\frac{3}{4}\tensor{R}{_{\mu}^{\nu}}\tensor{R}{_{\nu}^{\rho}}\tensor{R}{_{\rho}^{\sigma}}\tensor{R}{_{\sigma}^{\mu}}+\frac{5}{8}(R_{\mu\nu }R^{\mu\nu })^2+R\tensor{R}{_{\mu}^{\nu}}\tensor{R}{_{\nu}^{\rho}}\tensor{R}{_{\rho}^{\mu}}\\
&-\frac{7}{8}R^2R_{\mu\nu }R^{\mu\nu }+\frac{3}{32}R^4\, ,\\\notag
\mathcal{R}^{B}_{(4)}=&\frac{3}{16}R^{\mu\nu\rho\sigma }\tensor{R}{_{\mu\nu }^{\delta\gamma }}\tensor{R}{_{\delta\gamma }^{\chi\xi}}R_{\rho\sigma \chi\xi}-\frac{3}{64}(R_{\mu\nu\rho\sigma }R^{\mu\nu\rho\sigma })^2-\frac{1}{4}R\tensor{R}{_{\mu}^{\rho}_{\nu}^{\sigma}}\tensor{R}{_{\rho}^{\delta}_{\sigma}^{\gamma}}\tensor{R}{_{\delta}^{\mu}_{\gamma}^{\nu}}
-\frac{9}{4}R^{\mu\nu }R^{\rho\sigma }\tensor{R}{^{\delta\gamma }_{\mu\rho}}R_{\delta\gamma \nu \sigma}\\\notag
&-\frac{3}{2}R^{\mu\nu }\tensor{R}{_{\nu}^{\rho}}\tensor{R}{^{\sigma \delta\gamma }_{\mu}}R_{\sigma\delta\gamma \rho}+\frac{3}{16}R^2R_{\mu\nu\rho\sigma }R^{\mu\nu\rho\sigma}
+\frac{15}{4}RR_{\mu\nu\rho\sigma }R^{\mu\rho}R^{\nu\sigma}-\frac{15}{4}\tensor{R}{_{\mu}^{\nu}}\tensor{R}{_{\nu}^{\rho}}\tensor{R}{_{\rho}^{\sigma}}\tensor{R}{_{\sigma}^{\mu}}+\frac{21}{8}(R_{\mu\nu }R^{\mu\nu })^2\\
&+7R\tensor{R}{_{\mu}^{\nu}}\tensor{R}{_{\nu}^{\rho}}\tensor{R}{_{\rho}^{\mu}}-6R^2R_{\mu\nu }R^{\mu\nu }+\frac{47}{64}R^4\, ,\\\notag
\mathcal{R}^{C}_{(4)}=&-4R^{\mu\nu }\tensor{R}{_{\nu}^{\rho}}\tensor{R}{^{\sigma \delta\gamma }_{\mu}}R_{\sigma\delta\gamma \rho}+R_{\delta\gamma }R^{\delta\gamma }R_{\mu\nu\rho\sigma}R^{\mu\nu\rho\sigma }+4 RR_{\mu\nu\rho\sigma }R^{\mu\rho}R^{\nu\sigma}+\frac{16}{3}R\tensor{R}{_{\mu}^{\nu}}\tensor{R}{_{\nu}^{\rho}}\tensor{R}{_{\rho}^{\mu}}-5R^2R_{\mu\nu }R^{\mu\nu }\\
&+\frac{2}{3}R^4\, .
\end{align}
\end{widetext}

Out of these densities, only $\mathcal{R}_{(4)}^{A}$ contributes to the FL equations and it is normalized according to \req{eq:normR}. On the other hand, $\mathcal{R}_{(4)}^{B}$ contributes to the linearized equations around FLRW backgrounds but not to the FL equations, while $\mathcal{R}_{(4)}^{C}$ contributes to neither and is therefore irrelevant for cosmology. The full quartic term with the proper normalization can be written as a linear combination,
\begin{equation}\label{eq:quartic}
\mathcal{R}_{(4)}=\mathcal{R}_{(4)}^{A}+\nu \mathcal{R}_{(4)}^{B}+\gamma \mathcal{R}_{(4)}^{C}\, .
\end{equation}
where $\nu$ and $\gamma$ are free parameters. 

Higher-order densities satisfying conditions 1 and 2 were obtained in \cite{Arciniega:2018tnn} up to $n=8$, although the point 3 was overlooked. Nevertheless, given the present amount of evidence, we see no obstruction for the existence of densities satisfying points 1-3 at all orders in curvature. 

One interesting observation made in \cite{Arciniega:2018tnn} is that we do not need to know the explicit form of these Lagrangians in order to derive the FL equations. In fact, we already remarked that the current evidence allows one to conclude that at each order in the curvature these terms modify in a unique way the FL equations. By studying the first cases $n=3,\ldots 8$, one notices that the modification takes a simple form and that it is possible to derive a pattern for general $n$. On the other hand, we have checked that no simple pattern seems to exist for the perturbation equations and in fact at each order there may be more than one way in which these geometric terms modify the linearized equations --- see appendix \ref{app:tensor}. However, this shall be sufficient for our purposes since we will only carry out a detailed analysis of cosmological perturbations in the case of the cubic and quartic theory. 

\subsection{Geometric extensions of $R^2$ inflation}

The simplest version of Starobinsky's model of inflation is based on the following action, 
\begin{equation}\label{eq:SR2}
S=\frac{1}{16\pi G}\int d^4x\sqrt{|g|}\left[R+\frac{\alpha}{12}\ell^2R^2\right]\, ,
\end{equation}
where $\alpha>0$ is a dimensionless coupling constant that we take to be positive. Despite having fourth-order equations of motion, this theory is equivalent to a scalar-tensor theory with second-order EOMs and hence it is a viable cosmological model. 

We note that at the same order in the curvature expansion, we may add to the action the Gauss-Bonnet term $\mathcal{X}_{4}$ and the Weyl$^2$ term. The first one is topological, but the Weyl$^2$ term affects the tensor perturbations. It gives rise to fourth-order tensor perturbation equations and therefore ghosts, although these can plausibly be dealt with at least in certain situations \cite{Hawking:2001yt,Anselmi:2017ygm}. Also at the same order one may consider the effect of the trace anomaly of the matter fields stress-energy tensor on the Einstein's equations \cite{Hawking:2000bb} as Starobinsky originally did.  At higher orders in the derivative expansion many more terms can be added to the action, but for simplicity, we focus on the geometric terms we have described in the previous subsection. Since those terms give rise to second-order equations for the background evolution and for the linear perturbations, these provide an interesting and computationally feasible variation of the $R+R^2$ model. 

Thus, we consider the following purely metric model of inflation,
\begin{equation}\label{eq:SRn}
S=\frac{1}{16\pi G}\int d^4x\sqrt{|g|}\left[R+\frac{\alpha}{12}\ell^2R^2+\sum_{n=2}^{\infty} \lambda_n\ell^{2n-2}\mathcal{R}_{(n)}\right]
\end{equation}
in which the $R^2$ term is treated in a non-perturbative manner and the remaining terms as corrections.
As anticipated, due to the $R^2$ term, the Ricci scalar itself becomes a propagating degree of freedom in this theory. In order to make this manifest it is convenient to work with the following equivalent action instead:
\begin{align}\notag
S=\frac{1}{16\pi G}\int d^4x\sqrt{|g|}\Bigg[&R(1+2\alpha\ell^2\psi)-12\alpha \ell^2\psi^2\\
&+\sum_{n=2}^{\infty} \lambda_n\ell^{2n-2}\mathcal{R}_{(n)}\Bigg]\, .
\label{Spsi}
\end{align}
The equations of motion obtained from the variation of the metric and of the scalar field $\psi$ read, respectively
\begin{align}
\notag
\left(1+2\alpha\ell^2\psi\right)G_{\mu\nu}+\sum_{n=2}^{\infty} \lambda_n\ell^{2n-2}\mathcal{E}^{(n)}_{\mu\nu}&\\
\label{eq:Einstein}
- 2\alpha\ell^2\Big[\nabla_{\mu}\nabla_{\nu}\psi-g_{\mu\nu}(3\psi^2+\nabla^2\psi)\Big]&=0\\
R-12\psi&=0\, .
\label{eq:psieom}
\end{align}
Here $\mathcal{E}^{(n)}_{\mu\nu}$ denotes the contribution to the equation of motion obtained from the variation of the density $\mathcal{R}_{(n)}$:

\begin{align}
\mathcal{E}^{(n)}_{\mu\nu}&=\tensor{P}{^{(n)}_{(\mu}^{\rho \sigma \gamma}} R_{\nu) \rho \sigma \gamma} -\frac{1}{2}g_{\mu \nu} \mathcal{R}_{(n)}+2 \nabla^\sigma \nabla^\rho P^{(n)}_{\mu \sigma\nu\rho}\, ,
\end{align}
where
\begin{align}
P^{(n)}_{\alpha \beta \rho \gamma}&=\frac{ \partial \mathcal{R}_{(n)}}{\partial R^{\alpha \beta \rho \gamma}}\, .
\end{align}
From the second equation \req{eq:psieom} we get $\psi=R/12$, and plugging this value into the first equation we recover the equations of motion of the original theory \req{eq:SRn}.

\subsection{Generalized Friedmann-Lema\^itre equations} 

Consider a flat cosmological FLRW ansatz,
\begin{equation}
ds^2=-dt^2+a(t)^2\left(dx^2+dy^2+dz^2\right)\, .
\end{equation}
and $\psi(t)$. Inserting this in the EOM we obtain

\begin{align}
\label{eq:tt}
F\left(H^2\right)&=2\alpha\ell^2\left[\psi(\psi-H^2)-H\dot\psi\right]\, ,\\
\label{eq:xx}
\dot H F'\left(H^2\right)&=-\alpha\ell^2\left[\ddot\psi-H\dot\psi+2\dot H\psi\right]\, ,\\
\label{eq:psi}
6\dot H+12 H^2-12\psi&=0\, ,
\end{align}
where $H=\dot a/a$, with $\dot \psi=d\psi/dt$, etc, and $F$ is a function defined as
\begin{equation}\label{eq:Function}
F(x)=x+\ell^{-2}\sum_{n=3}^{\infty} (-1)^n\lambda_n \left(\ell^{2}x\right)^n\, .
\end{equation}
In addition, $F'(x)=dF/dx$. 
Note that in the absence of higher-order corrections we have $F(H^2)=H^2$, and the equations above reduce to those of $R^2$ gravity.  As we already remarked earlier, the densities $\mathcal{R}_{(n)}$ do not introduce higher derivatives in the equations of motion but modify Einstein's equations only through the function $F(H^2)$, which essentially determines the response of the geometry to the matter. In this case the ``matter'' is the inflaton field generated by the $R^2$ term. 

Note that the equation \req{eq:xx} can be derived from \req{eq:tt} and \req{eq:psi} by taking derivatives of those, and hence we do not need to include it in the resolution process. Thus, as usual, we can keep only the equations \req{eq:tt} and \req{eq:psi} which are of first order and define a dynamical system for $H$ and $\psi$. Since we shall be interested in inflationary solutions, it is useful to parameterize the evolution of the system in terms of the number of e-folds, $N=\log\left(a_0/a(t)\right)$, measured backwards from the end of inflation This is related to the time variable $t$ according to 
\begin{equation}
dN=-Hdt\, .
\end{equation}
Thus, from \req{eq:tt} and \req{eq:psi} we get the following system of equations,

\begin{align}\label{dpsidN}
\frac{d\psi}{dN}&=\frac{1}{H^2}\left[-\psi(\psi-H^2)+\frac{1}{2\alpha\ell^2}F(H^2)\right]\, ,\\
\label{dHdN}
\frac{dH}{dN}&=-\frac{2}{H}(\psi-H^2)\, .
\end{align}

Given that the origin of the $R^2$-term potentially differs from the other corrections we will often have in mind that $\alpha \gg 1$, while $\lambda_n\sim 1$. We can then distinguish three different regimes

\begin{enumerate}
	\item A general relativity regime when $1 \gg \alpha \ell^2 \psi \gtrsim \ell^2 H^2$ 
	
	\item An $R^2$-phase when $\alpha \ell^2 \psi  \gtrsim 1 \gg \ell^2 H^2$ 
	
	\item A higher-derivative phase when $\alpha \ell^2 \psi  \gtrsim  \ell^2 H^2 \gtrsim 1$ 
\end{enumerate}
Thus, we expect that the higher-order terms are especially relevant earlier on during inflation. Since our theories depend on the choice of an arbitrary function $F(H^2)$, in the following we shall first investigate the choices of this $F$ that lead to desirable cosmologies. By this we mean that

\begin{enumerate}	
	\item the theory admits a sufficiently long inflationary regime.
	
	\item Inflation acts as an attractor in the sense that there is an open set of initial conditions that give rise to inflation. 
	
	\item Inflation ends in the general relativity regime.
\end{enumerate}

First, however, it should be remarked that when there are no higher derivative corrections in addition to $R^2$, these conditions are satisfied with an appropriate choice of $\alpha$ and/or the characteristic scale $\ell$. 
Taking $R^2$ inflation as our reference model, we are interested in the range of $\lambda_n$ that preserve these nice properties. One particular point that should be considered is the initial value of the scalar $\psi$ (equivalently $R$) required to produce a sufficiently large number of e-folds. The higher-order corrections will modify this value with respect to the $R^2$ prediction.

\section{slow roll regime}
\label{sec:slowroll}
Before studying the full dynamical system given by \req{dpsidN} and \req{dHdN}, we analyze its slow roll regime. A necessary condition in order for a theory to be a viable candidate model of inflation is that its slow roll regime allows for sufficiently many e-folds\footnote{It should be noted that, to match cosmological observations, the required number of such e-folds is model-dependent. In particular, it increases with the reduction of curvature scales from $\psi_{\rm obs}$ to $\psi_{\rm end}$ as  $\sim\frac{1}{2}\log{\frac{\psi_{\rm obs}}{\psi_{\rm end}}}$, which could become significant in a subset of the theories under consideration. However, we will later derive holographic constraints which effectively rule out that possibility.} $N$. Typically one needs $N$ in the range of $50$ to $60$, reheating being the largest source of uncertainty \cite{liddle:2003as}.

It is customary to introduce the slow roll parameter $\epsilon$,
\begin{equation}
\epsilon=\frac{1}{H}\frac{dH}{dN}\, ,
\end{equation}
so that, during inflation, $\epsilon\ll1$, and inflation ends when $\epsilon\approx 1$. From \req{dHdN} we obtain 
\begin{equation}
\epsilon=-\frac{2}{H^2}(\psi-H^2)\, .
\end{equation}

Replacing $H$ in terms of $\psi$ and $\epsilon$ in \req{dpsidN} yields
\begin{equation}
\frac{d\psi}{dN}=\frac{1}{2}\epsilon\psi+\frac{1-\epsilon/2}{2\alpha\ell^2\psi}F\left(\frac{\psi}{1-\epsilon/2}\right)\, .
\end{equation}
Now, let us take into account that, since $H^2\approx\psi$ during inflation, we have the following approximation
\begin{equation}
\epsilon\approx \frac{1}{2\psi}\frac{d\psi}{dN}\, .
\end{equation}
Therefore, to leading order in $\epsilon$:
\begin{equation}\label{slowroll1}
\frac{d\psi}{dN}\approx \frac{2F(\psi)}{3\alpha\ell^2\psi}\, .
\end{equation}
This equation governs the dynamics of the system during the slow roll phase. On the other hand, $\epsilon$ is given in terms of $\psi$ by
\begin{equation}\label{slowroll2}
\epsilon\approx  \frac{F(\psi)}{3\alpha\ell^2\psi^2}\, .
\end{equation}

\begin{figure}[t!]
	\begin{center}
		\includegraphics[width=0.49\textwidth]{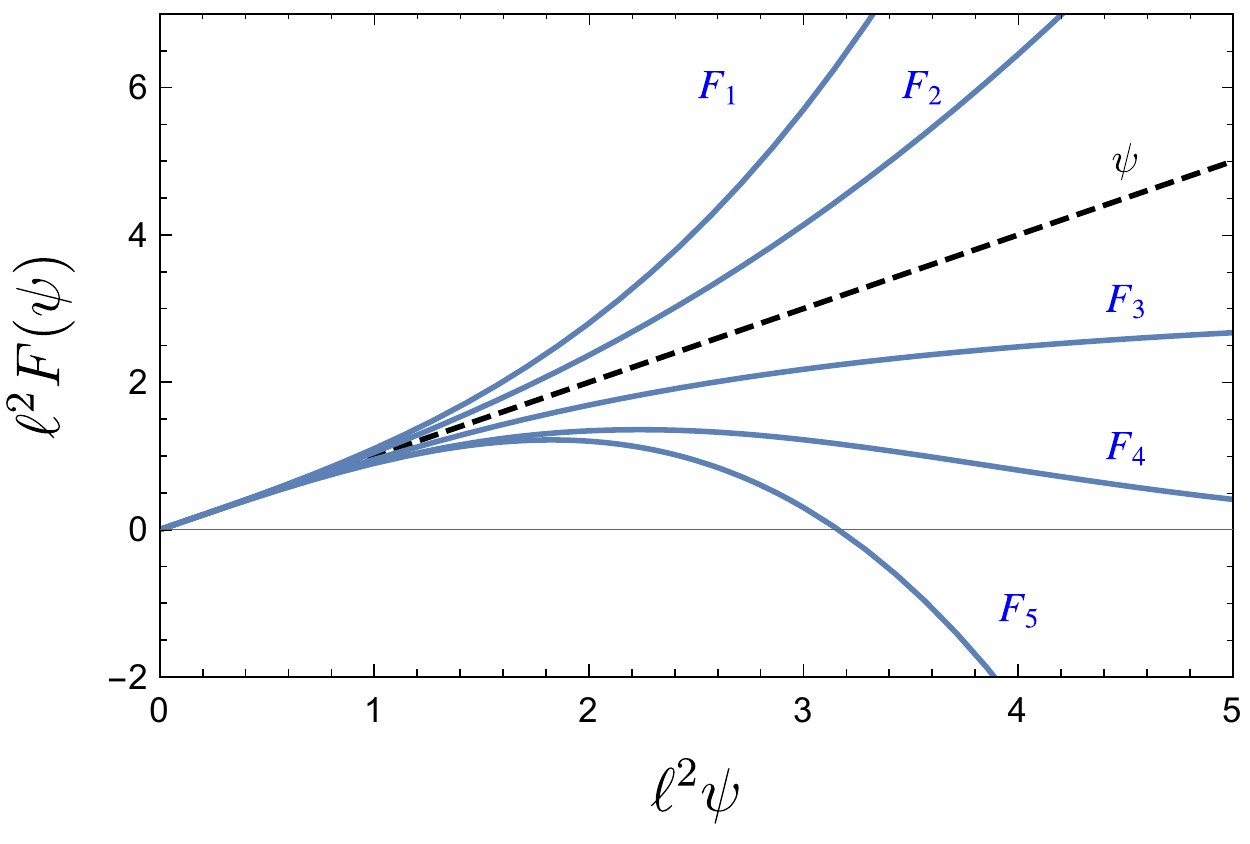} 
		\caption{Representative examples of the function $F(\psi)$ that characterizes the higher derivative gravity at the background level. The expressions for $F_1$, $F_2$, $F_3$, $F_4$ and $F_5$ are given respectively in \req{eqn:F1}, \req{eqn:F2}, \req{eqn:F3}, \req{eqn:F4} and \req{eqn:F5}, each with $\lambda=0.1$. As a reference, the GR case $F(\psi) = \psi$ is also shown.}
		\label{fig:Ffunctions}
	\end{center}
\end{figure}

\begin{figure*}[htpb]
	\begin{center}
		\includegraphics[width=0.49\textwidth]{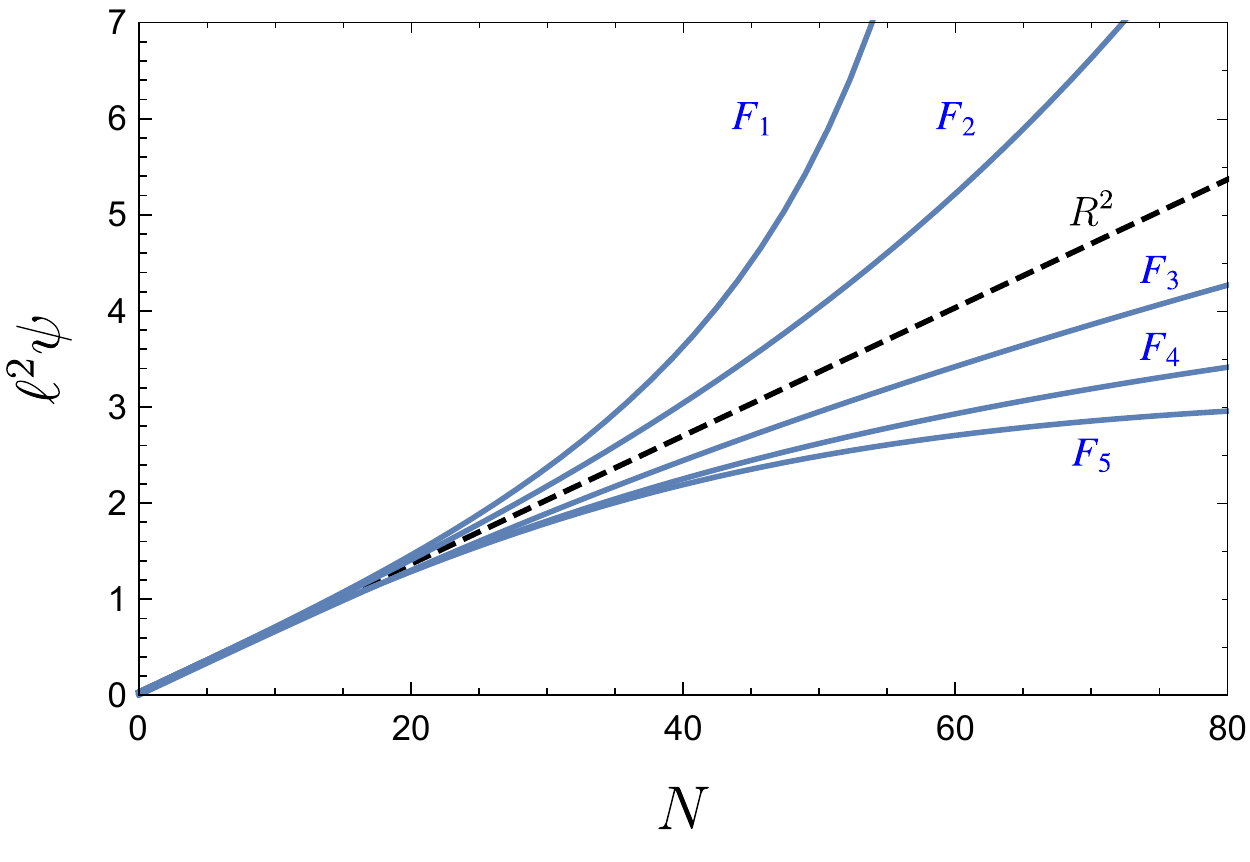} 
		\includegraphics[width=0.49\textwidth]{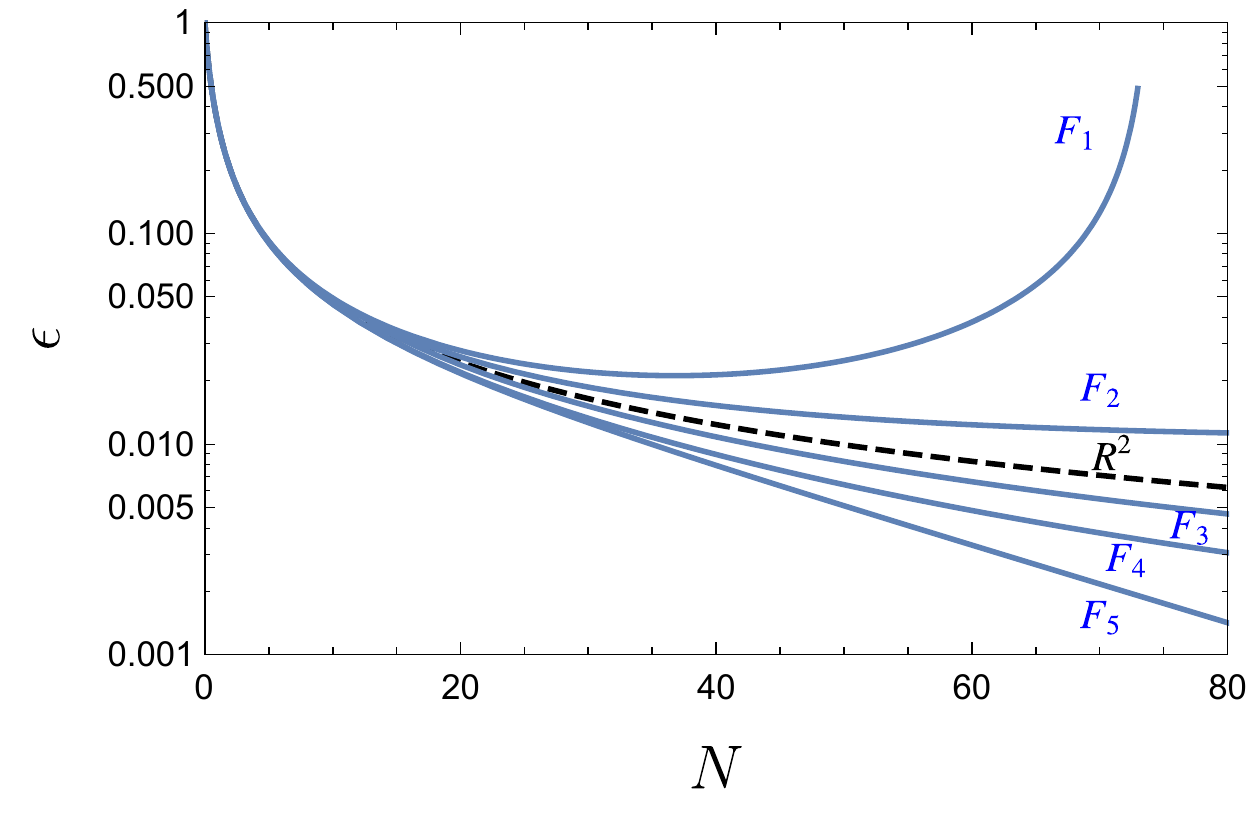}
		\caption{Slow roll evolution of the curvature (left) and slow roll parameter (right) for the different higher-derivative theories discussed in Sec. \ref{sec:slowroll}. The $F(\psi)$ functions that determine the examples at the background level are shown in Fig.~\ref{fig:Ffunctions}. The parameters used in each case are $\alpha=10$, $\lambda=0.1$. The approximate expressions for these slow roll evolutions are given in \req{eqn:F1psi}, \req{eqn:F2psi}, \req{eqn:F3psi} (asymptotically), \req{eqn:F4psi} and \req{eqn:F5psi}.}
		\label{fig:Fcomparison}
	\end{center}
\end{figure*}

Our goal in this section is to determine which conditions $F$ should satisfy in order for there to be a large ($N\sim 60$) number of e-folds. 
For $R^2$ inflation without additional higher-order terms we have $F(\psi)=\psi$, and the equation \req{slowroll1} can be integrated to yield
\begin{equation}
\psi=\psi_{\rm end}+\frac{2N}{3\alpha\ell^2}\, .
\end{equation}
We recall that $N$ measures the e-folds till the end of inflation.  On the other hand, $\epsilon=(3\alpha\ell^2\psi)^{-1}$, so that inflation ends at $\psi_{\rm end}\sim (3\alpha\ell^2)^{-1}$, which implies that
\begin{equation}\label{eqn:slowpsiR2}
\psi=\frac{1+2N}{3\alpha\ell^2}\, ,\quad \epsilon=\frac{1}{1+2N}\, .
\end{equation}
Notice that the initial value $\psi_{\rm ini}=\frac{1+2N_{\rm total}}{3\alpha\ell^2}$ can be quite large in units of $\ell^{-2}$, since $N_{\rm total}$ should be at least $\sim 60$. Therefore, the possible higher-order effects can be very relevant. 
Of course, due to the freedom in choosing the function $F$, the possible number of situations is endless. We distinguish three general cases that we consider most natural. A simple possibility is to assume that the higher-order corrections modify the function $F$ in a monotonous manner, so that it either grows faster than in $R^2$ inflation or more slowly. A third situation arises when $F$ has a zero. As an overview of the following discussion, Fig.~\ref{fig:Ffunctions} shows the representative examples we will consider in each case while Fig.~\ref{fig:Fcomparison} gives a comparison between the evolution of $\psi$ and $\epsilon$ in the different cases.

\subsection{$F$ is larger than in GR: $F(\psi)>\psi$}
First assume that $F$ is modified by the higher-order corrections so that it is larger than in GR. We see from Eq.~\req{slowroll1} that this implies that the velocity at which the scalar field ``rolls down'' will be increased, meaning that the initial value in order to achieve 60 e-folds will be higher. On the other hand, the slow roll parameter \req{slowroll2} becomes larger than in $R^2$ inflation. Thus, in this case the higher-order corrections tend to spoil the nice properties of the $R^2$ scenario, although inflation is still possible depending on the values of the couplings. It is convenient to consider two subcases that we illustrate next with two examples.

\subsubsection{Asymptotically $F(\psi)>\psi^2$}
The first subcase corresponds to having a function $F(\psi)>\psi$ and such that $F(\psi)>\psi^2$ when $\psi\rightarrow\infty$. A very simple example is given by the choice of parameters $\lambda_3=-\lambda>0$, $\lambda_{n>3}=0$, so that
\begin{equation}
F_{1}(\psi)=\psi+\lambda \ell^4\psi^3\, ,
\label{eqn:F1}
\end{equation}
which corresponds to simply having a cubic term in the action.  Now, integrating \req{slowroll1} yields
\begin{equation}
\frac{1}{\sqrt{\lambda}}\arctan(\sqrt{\lambda}\ell^2\psi)-\frac{1}{\sqrt{\lambda}}\arctan(\sqrt{\lambda}\ell^2\psi_{\rm end})=\frac{2N}{3\alpha}\, .
\end{equation}
On the other hand, the slow roll parameter reads

\begin{equation}
\epsilon=\frac{1+\lambda\ell^4\psi^2}{3\alpha\ell^2\psi}\, .
\end{equation}
At the end of inflation $\epsilon=1$, so that

\begin{equation}
\ell^2 \psi_{\rm end}=\frac{1}{2 \lambda }\left[3\alpha-\sqrt{9\alpha^2-4\lambda}\right]\, ,
\end{equation}

and therefore

\begin{align}
\ell^2\psi=\frac{1}{\sqrt{\lambda}}\tan\left(\frac{2\sqrt{\lambda}N}{3\alpha}+ \arctan\left(\frac{3\alpha }{2 \sqrt{\lambda}}-\sqrt{\frac{9\alpha^2}{4 \lambda}-1}\right)\right)\, .
\label{eqn:F1psi}
\end{align}
As we can see, a peculiarity of this model is that $\psi$ diverges at finite $N$ --- a feature shared by any other model with $F(\psi)>\psi^2$ asymptotically. In other words, this implies that there is a maximum amount of e-folds, which is approximately
\begin{equation}\label{eq:F1max}
N_{\rm max}\approx \frac{3\pi\alpha}{4\sqrt{\lambda}}\, ,
\end{equation}
when $\sqrt{\lambda}<<\alpha$. Thus, the ratio $\lambda/\alpha^2$ must be small enough in order to achieve inflation. 

\subsubsection{Asymptotically $F(\psi)\le\psi^2$}

A second subcase of $F(\psi) > \psi$ has asymptotically $F(\psi)\le\psi^2$. Contrary to the previous subcase, the curvature only diverges asymptotically towards the past. This behavior can be illustrated by choosing, for example 
\begin{equation}
F_{2}(\psi)=\psi\sqrt{1+\lambda\ell^4\psi^2}\, ,\quad \lambda>0 \, .
\label{eqn:F2}
\end{equation}
Note that in this case would correspond to a summation of \req{eq:Function} for a particular choice of the infinite set of higher-order couplings $\{\lambda_{n}\}_{n\ge3}$. In the slow roll regime we have

\begin{equation}
\epsilon = \frac{\sqrt{1+\lambda\ell^4\psi^2}}{3\alpha \ell^2 \psi}\, ,
\end{equation}

and

\begin{equation}
\ell^2  \psi = \frac{1}{\sqrt{\lambda}}\sinh{\left( \frac{2 \sqrt{\lambda} N}{3 \alpha} + \rm \operatorname{arcsinh}\left(\frac{\sqrt{\lambda}}{\sqrt{9\alpha^2-\lambda}}\right) \right)}\, .
\label{eqn:F2psi}
\end{equation}
In this case it is possible to achieve an arbitrary number of e-folds, although this may require a quite large initial value of $\psi$.

\subsection{$F$ is smaller than in GR: $0<F(\psi)<\psi$}

Assuming instead that $F$ is modified by the higher order corrections to be smaller than in GR, $0<F(\psi)<\psi$, more e-folds will be realized given similar initial values. Such higher-curvature corrections would therefore be theoretically preferable if inflation is to happen at modest relative curvatures or for small $\alpha$. Let us again consider two subcases with representative examples.

\subsubsection{Asymptotically $F(\psi) = O(\psi^{\alpha})$ for some $\alpha<1$}

This first subcase of $0<F(\psi)<\psi$, has $F(\psi)$ behaving as a power law for $\psi\rightarrow\infty$. Consider the following example

\begin{equation}
F_{3}(\psi)=\frac{\psi}{\sqrt{1+\lambda\ell^4\psi^2}} ,\quad \lambda>0 \, ,
\label{eqn:F3}
\end{equation}
which is a monotonous function that tends to a constant value for $\psi\rightarrow\infty$. By Eq.~\req{slowroll1}, the slow roll evolution satisfies

\begin{equation}
\ell^2 \psi\sqrt{1+\ell^4  \psi^2 \lambda}+
\frac{1}{\sqrt{\lambda}}\operatorname{arcsinh}(\ell^2  \psi \sqrt{\lambda}) =  \frac{4 (N-N_0)}{3 \alpha}\, ,
\end{equation}

with 

\begin{align}
-\frac{4N_0}{3\alpha}&=\ell^2 \psi_{\rm end}\sqrt{1+\ell^4  \psi_{\rm end}^2 \lambda}+
\frac{1}{\sqrt{\lambda}}\operatorname{arcsinh}(\ell^2  \psi_{\rm end} \sqrt{\lambda})\, , \\
\ell^2 \psi_{\rm end} &= \sqrt{\frac{-3+\sqrt{9+\frac{4\lambda}{\alpha^2}}}{6 \lambda}}\, .
\end{align}
In the limit of large $N$ one gets 
\begin{equation}\label{eqn:F3psi}
\ell^2\psi\approx\sqrt{\frac{4N}{3\alpha\sqrt{\lambda}}}\, ,
\end{equation}
so that $\psi$ scales with $N^{1/2}$ rather than linearly with $N$, and hence it is much smaller than in the $R^2$ case. In addition, the slow-roll parameter $\epsilon\approx 1/(4N)$ is half of the $R^2$ prediction. 

\subsubsection{Asymptotically $F(\psi) = o(\psi^{\alpha})$ for all $\alpha<0$}

The second subcase of $0<F(\psi)<\psi$, has $F$ going to zero faster than any inverse power law. For instance

\begin{equation}
F_{4}(\psi)=\psi e^{-\lambda\ell^4\psi^2} ,\quad \lambda>0
\label{eqn:F4}\, .
\end{equation}

The slow roll curvature evolution of this model is given by

\begin{equation}
\ell^2  \psi = \frac{1}{\sqrt{\lambda}} \operatorname{erfi}^{-1}\left(  \frac{4 \sqrt{\lambda} N}{3 \alpha \sqrt{\pi}} + \operatorname{erfi}\left( \sqrt{W\left(\frac{2 \lambda}{9 \alpha^2}\right)/2}\right)\right)\, ,
\label{eqn:F4psi}
\end{equation}
with $\rm erfi$ the imaginary error function and $W$ Lambert's productlog function. In this case, $\psi$ depends logarithmically on $N$ in the limit of large $N$, so one can achieve a very long inflationary regime with relatively small values of the curvature. 

\subsection{$F$ has a zero for positive $\psi$}

In case $F$ has a zero for positive $\psi$, there is a fixed point corresponding to an exact de Sitter vacuum. This can easily give rise to a long period of inflation, the danger now rather being that the GR regime is never reached. For example, take $F_1$ with $\lambda \to -\lambda$

\begin{equation}
F_{5}(\psi)=\psi-\lambda \ell^4\psi^3\, ,\quad \lambda>0\, ,
\label{eqn:F5}
\end{equation}
so that the slow-roll evolution is equivalent to Eq. \req{eqn:F1} with $\lambda \to - \lambda$,

\begin{align}
\ell^2\psi=\frac{1}{\sqrt{\lambda}}\tanh\left(\frac{2\sqrt{\lambda}N}{3\alpha}+ \operatorname{arctanh}\left(\sqrt{\lambda}\psi_{\rm end}\right)\right)
\label{eqn:F5psi}\, .
\end{align}
This can be considered to have evolved from the fixed point $\ell^2\psi = \frac{1}{\sqrt{\lambda}}$ asymptotically to the past and in this way is in practice not very different from the example in Eq.~\req{eqn:F4} above. However, it is not the only possibility as one can also be driven away from the fixed point with growing curvature if, in this case, $\ell^2\psi > \frac{1}{\sqrt{\lambda}}$. In that situation the slow roll parameter is negative

\begin{equation}
\epsilon = \frac{1-\lambda \ell^4 \psi^2}{3 \alpha l^2 \psi}\, ,
\end{equation}
meaning that inflation accelerates.  The slow-roll regime ends around $\epsilon = -1$, for which

\begin{equation}
\ell^2 \psi_{\rm end} = \frac{1}{2 \lambda}(3\alpha + \sqrt{9 \alpha +4 \lambda})\,,
\end{equation}
but at this point the solution keeps expanding at an accelerated rate and the GR regime as never reached.  Clearly this is undesirable. The behavior around general fixed points will be analyzed in more detail in the next section where we explore the phase space of \req{dpsidN} and \req{dHdN} beyond the slow roll regime.
 
\section{Attractor mechanism}\label{sec:attractor}

Our analysis in the previous section was restricted to the slow roll regime of the cosmologies. We now explore the full phase space of the system \req{dpsidN} - \req{dHdN} and show that in a wide range of the higher-curvature extensions of $R^2$ inflation, slow roll inflation is an attractor solution. We also derive a convenient bound, Eq.~\req{eqn:HDbound}, that quantifies the initial curvature scale that is required to obtain a prolonged phase of inflation that subsequently transitions into a FLRW regime governed by GR. 

First, note that aside from the term $\frac{1}{2\alpha\ell^2}F(H^2)$, the dynamical system \req{dpsidN}, \req{dHdN} tends towards $\psi \approx H^2$ for decreasing $N$, on the assumption that $\psi > 0$. To see this, combine \req{dpsidN} and \req{dHdN}  to 
\begin{equation}
-\frac{d}{d N}\left(\frac{\psi}{H^2}\right) = -  \frac{3 \psi}{H^2}\left(\frac{\psi}{H^2}-1\right)-\frac{F(H^2)}{2 \alpha \ell^2 H^4} \, .
\label{eqn:psiH2ratio}
\end{equation}
If $\frac{\psi}{H^2}(\frac{\psi}{H^2}-1) \gg \frac{F(H^2)}{2 \alpha \ell^2 H^4}$, the ratio of \req{dpsidN} and \req{dHdN} approximately becomes
\begin{equation}
\frac{d \psi}{d H} \approx \frac{\psi}{2 H} \, ,
\end{equation} 
which implies that $\psi \propto \sqrt{H}$. Now, inflation starts when $\psi\approx H^2$, and at that point we demand that $\psi\ge\psi_{\rm start}$, where  $\psi_{\rm start}$ is the minimum initial value of $\psi$ in order to have a sufficient number of e-folds $N_{\rm start}$. This translates into the following condition for the initial values $H_0$, $\psi_0$,
\begin{equation}
\ell^2 \psi_0 \geq (\ell^2 \psi_{\rm start})^{3/4} \sqrt{\ell H_0} \, ,
\label{eqn:HDbound}
\end{equation}
Of course this analysis neglects the contribution of the $\frac{F(H^2)}{2 \alpha \ell^2 H^4}$ term in Eq.~\req{eqn:psiH2ratio}, but this only becomes important when one approaches $\psi \approx H^2$. 

\begin{figure*}[htpb]
	\begin{center}
		\includegraphics[width=0.49\textwidth]{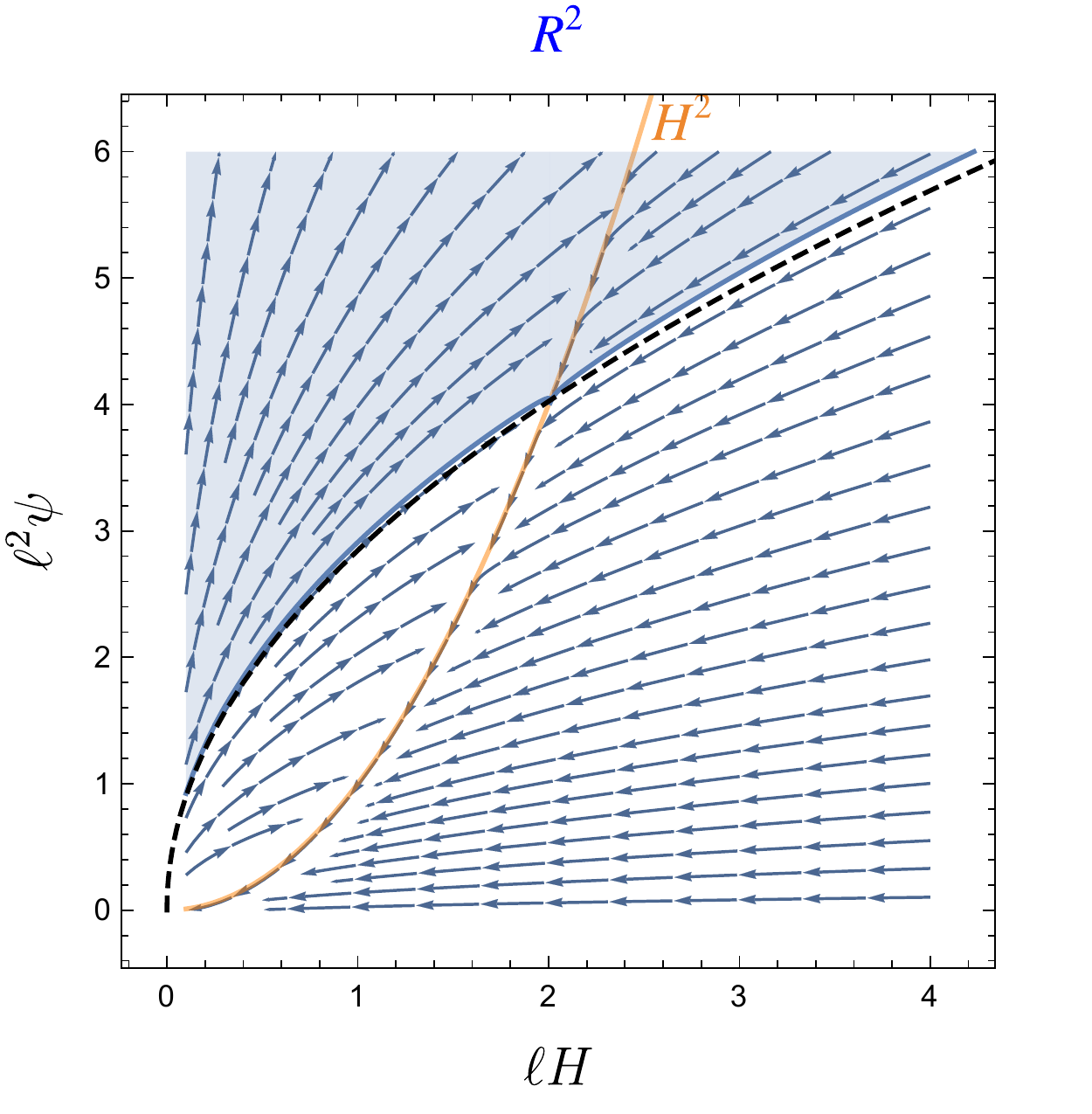} 
		\includegraphics[width=0.49\textwidth]{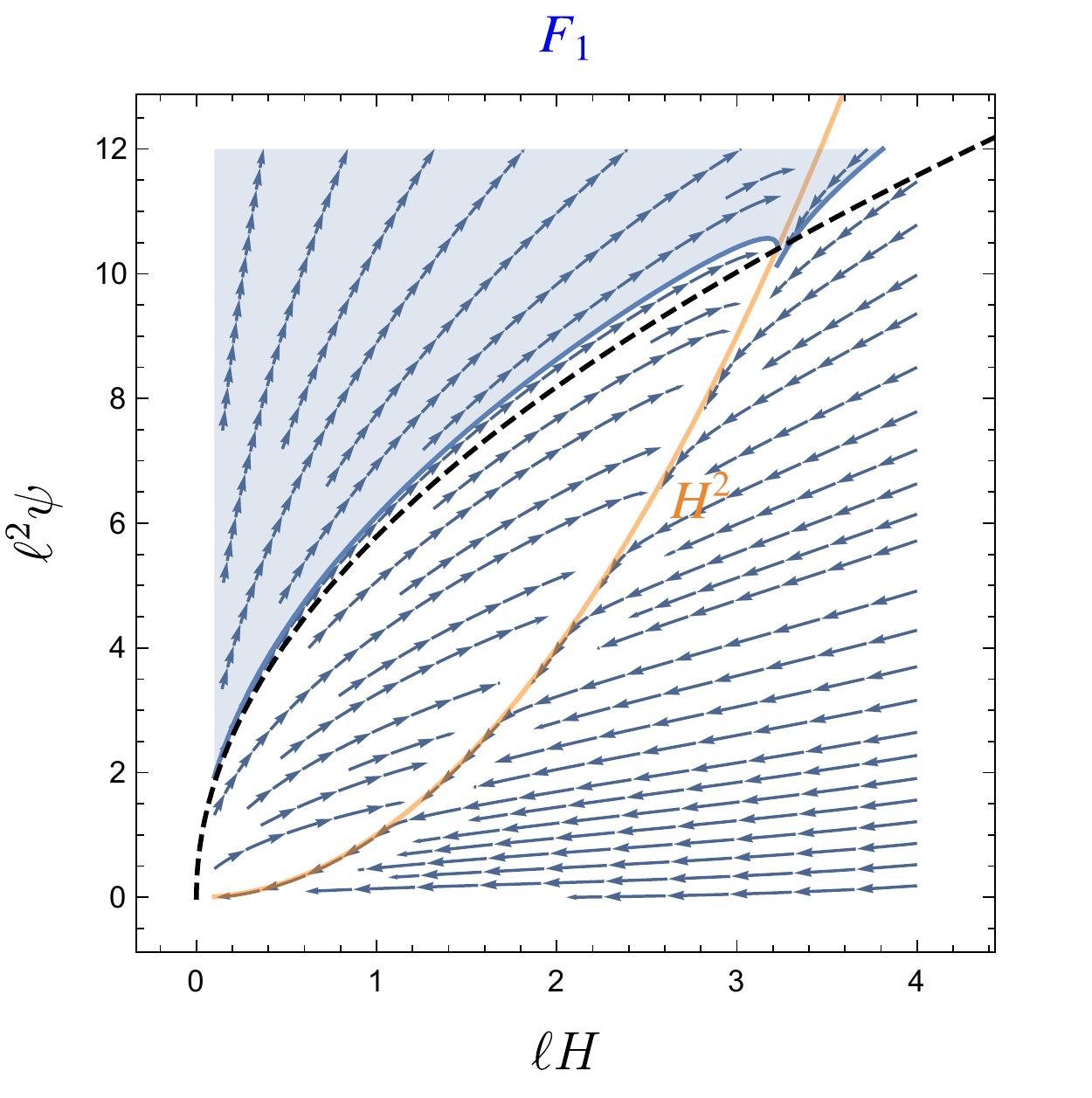} 
		\includegraphics[width=0.49\textwidth]{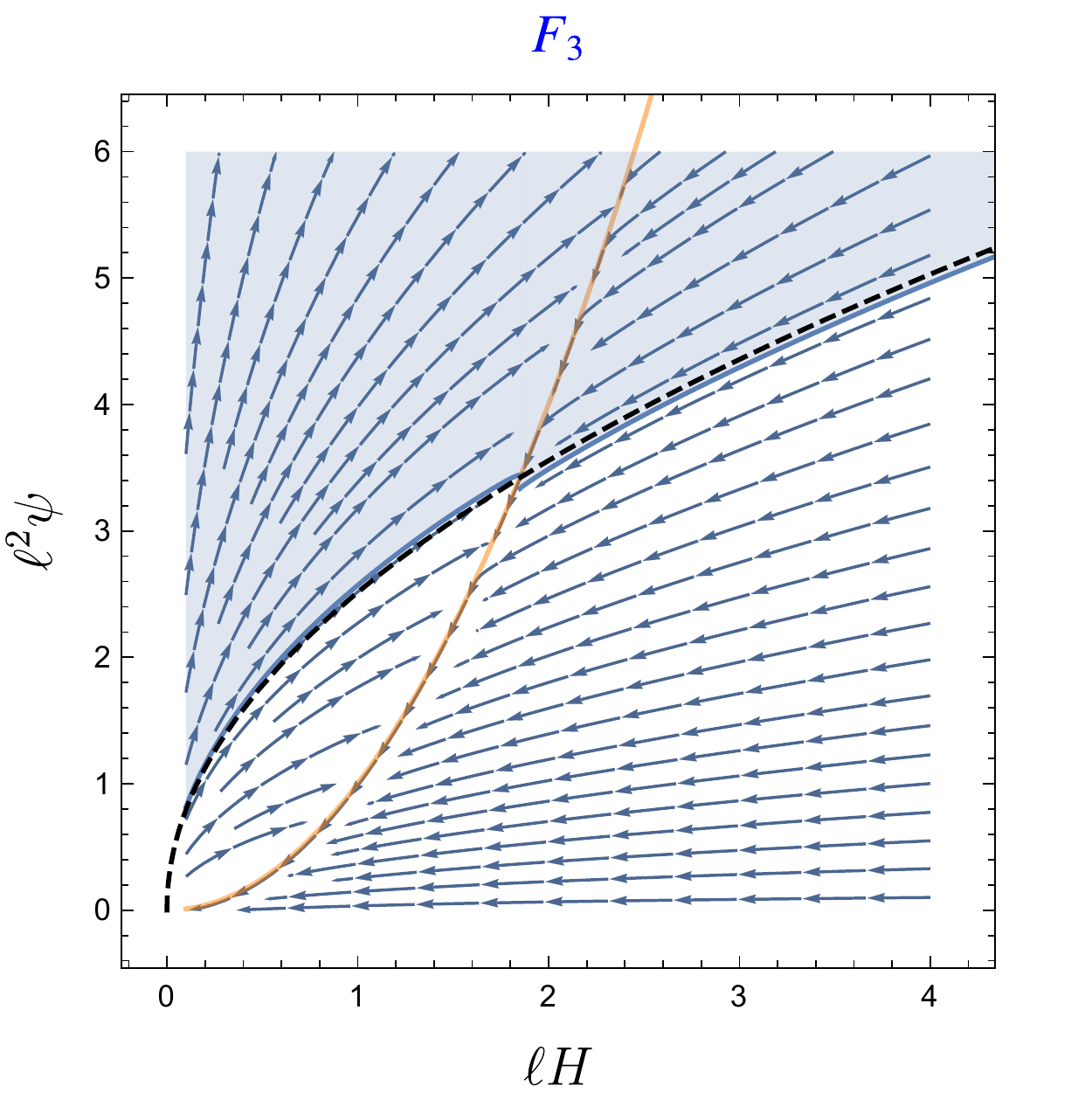} 
		\includegraphics[width=0.49\textwidth]{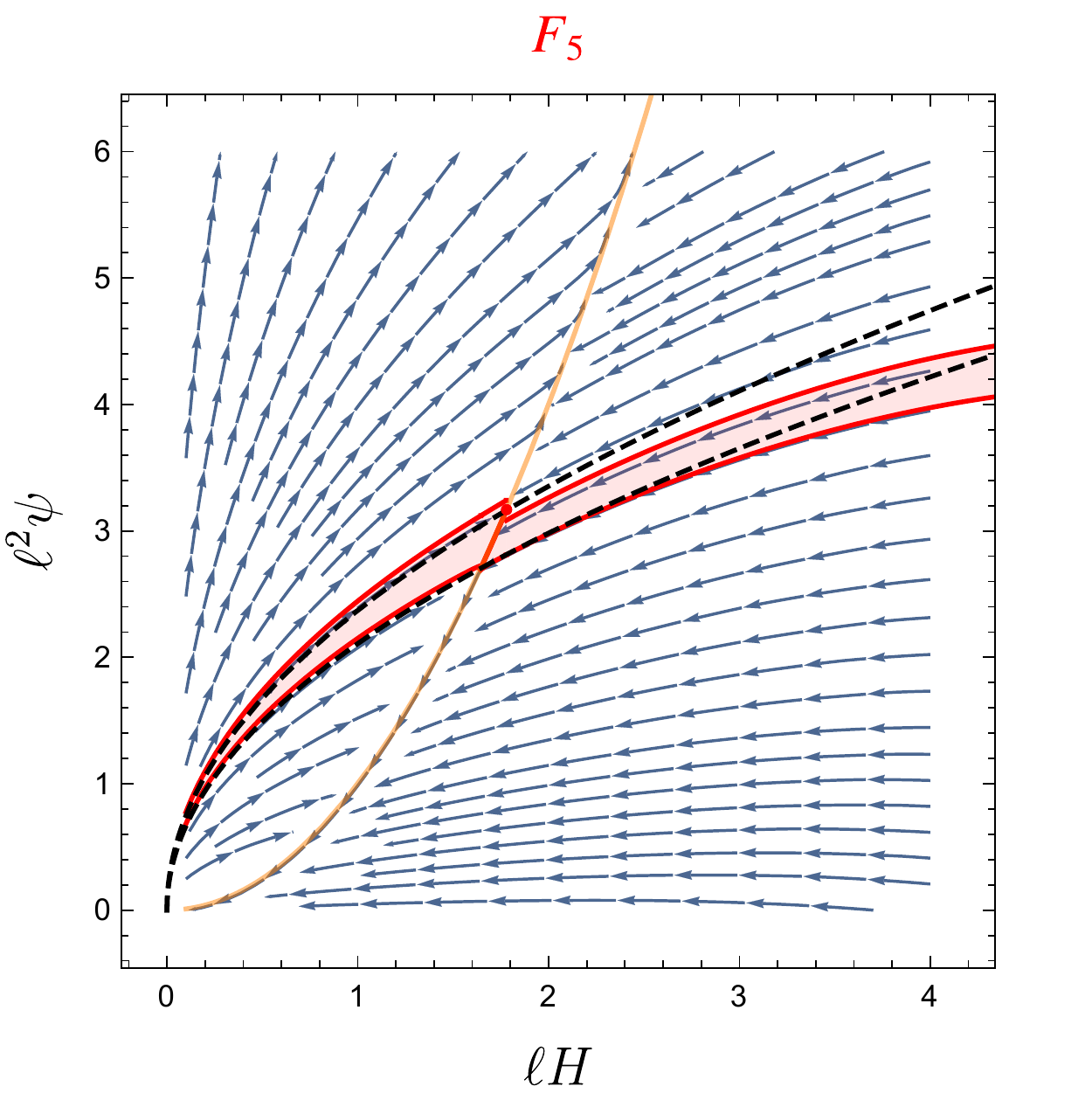} 
		\caption{Flow in phase space for the dynamical system \req{dpsidN}, \req{dHdN} in the case of pure $R^2$ (top left), $F_1$ as defined in Eq.~\req{eqn:F1} (top right), $F_3$ as defined in Eq.~\req{eqn:F3} (bottom left) and $F_5$ as defined in Eq.~\req{eqn:F5} (bottom right), for $\alpha=10$ and $\lambda=0.1$. The flow shows the evolution forward in time and the vector field is indicative of the direction of the flow, not its speed. Shown as well is the line $\psi =H^2$ (full, orange), indicative of the slow roll regime, and the region of phase space which gives rise $60$ e-folds of slow roll inflation (shaded) as well as the approximation Eq.~\req{eqn:HDbound} to the boundary of this region (dashed, black). For the model $F_5$ (bottom right), the additional nontrivial requirement that inflation ends in a regime described by general relativity leads to an upper bound, given approximately by \req{eqn:dSbound}, and strongly reduces the phase space region with a capacity to realise enough inflation.}
		\label{fig:R2phaseplot}
	\end{center}
\end{figure*}

Now consider instead $\frac{\psi}{H^2} \ll 1$. If $F(H^2) > 0$ and $\frac{\psi}{H^2} \ll \frac{F(H^2)}{2 \alpha \ell^2 H^4}$, then Eq.~\req{eqn:psiH2ratio} implies that $\frac{\psi}{H^2}$ will decrease further, away from possible slow roll. The only way to nevertheless reach slow roll would therefore be for $F(H^2)$ to decrease even faster
\begin{equation}
F(H^2) < \frac{4 H^2 F'(H^2)}{9} \, .
\end{equation} 
For $R^2$ this inequality does not hold, so we may just assume that it does not hold for the higher-order gravities either. Therefore, from the onset of inflation one must have $\frac{\psi}{H^2} \gtrsim \frac{F(H^2)}{2 \alpha \ell^2 H^4}$. For simplicity, say $\frac{\psi_0}{H^2_0} \gg \frac{F(H_0^2)}{2 \alpha \ell^2 H_0^4}$. Then we can apply the same reasoning as before and we are left again with the bound \req{eqn:HDbound}. 

For $R^2$ inflation, Eq.~\req{eqn:slowpsiR2} indicates that we need $\psi_{\rm start} = \frac{1+2N_{\rm start}}{3 \alpha \ell^2}$ to have $N_{\rm start}$ e-folds of slow roll inflation, while for the models introduced in Sec.~\ref{sec:slowroll}, $\psi_{\rm start}$ is (approximately) given by \req{eqn:F1psi}, \req{eqn:F2psi}, \req{eqn:F3psi}, \req{eqn:F4psi}, and \req{eqn:F5psi} as a function of the desired number of e-folds. For some of these theories, Fig.~\ref{fig:R2phaseplot} shows the flow in phase space and the region of initial conditions leading to  $N_{\rm start}>60$ e-folds of inflation. As we can see, the bound \req{eqn:HDbound} provides a reasonably good approximation for the allowed region of initial conditions, and, overall, it captures the main differences between these models.
Of course it should be noted that the differences become more pronounced as the contribution of $F(H^2)$ increases. Also, the approximation breaks down at large $H$.  

The fact that these theories give rise to a regular dynamical system with a global attractor towards inflation is in itself a non-trivial result. 
If one considered generic higher-order corrections to $R^2$ inflation, one would find singularities in the dynamical system and that inflation is not an attractor for arbitrary initial conditions. This can be checked for simple examples, such as adding a Riemann$^3$ term in the action rather than our cubic density \req{eq:cECG}. This is yet another reason why the set of theories we are considering is appealing for cosmology.

When the model has a fixed point and thus a sign change of $F(H^2)$, one needs to worry about the additional requirement that inflation ends in a phase described by general relativity. Consider the general case where $F(H^2)$ has a zero $H_{\rm dS}$. The dynamical system \req{dpsidN} - \req{dHdN} linearized about the fixed point $H = H_{\rm dS}$, $\psi = H_{\rm dS}^2$ is given by
\begin{equation}
-\begin{pmatrix} \frac{d \delta \psi}{d N} \\ \frac{d \delta H}{d N} \end{pmatrix} = \begin{pmatrix} 1 & -2H_{\rm dS} - \frac{F'(H_{\rm dS}^2)}{ H_{\rm dS} \ell^2 \alpha} \\ \frac{2}{H_{\rm dS}} & -4 \end{pmatrix} \begin{pmatrix} \delta \psi \\ \delta H \end{pmatrix}\, ,
\label{eqn:lineom}
\end{equation} 
with $\delta H = H-H_{\rm dS}$ and $\delta \psi = \psi-H^2_{\rm dS}$. The eigenvalues
\begin{equation}
\lambda_{\pm}=	-\frac{3}{2}\pm \sqrt{\frac{9}{4}-\frac{2 F'(H_{\rm dS}^2)}{H^2_{\rm dS} \ell^{2} \alpha}}\, ,
\end{equation}
and associated eigenvectors
\begin{equation}
e_{\pm} = \begin{pmatrix} \frac{H_0}{4} (5 \pm 2\sqrt{\frac{9}{4}-\frac{2 F'(H_{\rm dS}^2)}{H^2_{\rm dS} \ell^2 \alpha}}) \\ 1 \end{pmatrix}\, ,
\end{equation}
indicate that, when $F'(H_{\rm dS}^2) > 0$ the fixed point is stable\footnote{For increasing time or decreasing $N$.}, whereas when $F'(H_{\rm dS}^2) < 0$, the fixed point is a saddle. The latter case is most relevant since it occurs for the smallest positive zero in $F(H^2)$, assuming $F(H^2) \to H^2$ for small $H$ and this is what happens for the model $F_5$ \req{eqn:F5}. The saddle, in this case, is roughly attractive towards the line $\psi = H^2$ and unstable along this line. Crucially, when $\psi \approx H^2$ and $H > H_{dS}$, the evolution is now driven towards higher $H$, never to reach the general relativity regime $\ell H \ll 1$. This is shown, for the model \req{eqn:F5} in Fig.~\ref{fig:R2phaseplot} (bottom right) and it is the reason an upper bound must be introduced in this model. Following the same procedure as for Eq.~\req{eqn:HDbound}, this upper bound can be approximated as
\begin{equation}
\ell^2 \psi_0 \leq (\ell^2 \psi_{\rm dS})^{3/4} \sqrt{\ell H_0}\, ,
\label{eqn:dSbound}
\end{equation}
As Fig.~\ref{fig:R2phaseplot} shows, in this case there is only a narrow region in the phase space leading to successful inflation, and this region becomes narrower if we increase $\lambda$. Thus, if $F(\psi)$ has a zero we may need fine-tuned initial conditions in order to produce inflation. By contrast, functions $0<F(\psi)<\psi$ enlarge the region of initial conditions in the phase space leading to a long inflationary period.

\section{Hartle-Hawking Initial conditions}

We have seen that in a wide range of the geometric higher-curvature corrections to $R^2$ inflation that we consider, inflation is an attractor. This means there is an open set of initial values of $H$ and $\psi$ that give rise to inflation. This does not mean, however, that inflation is `likely'. Reasonable classical measures on phase space clearly point towards the need for a quantum theory of initial conditions to explain the origin of inflation \cite{Gibbons:2006pa}. Here we generalize earlier work \cite{Vilenkin:1985md, Hawking:2000bb} on the quantum cosmology of Starobinsky inflation to include the geometric higher-curvature corrections in the action.\footnote{See \cite{Jonas:2020pos} for a related analysis.} Second, we generalize the analysis of \cite{Hawking:2000bb} to so-called fuzzy instantons which describe not a sharp but a gradual transition from a quantum regime to a classical, Lorentzian slow roll inflationary phase. We will find that in any given model, the no-boundary wave function selects a one-parameter family of cosmologies which can be labelled by different starting values $\psi_0$ of the curvature.

\begin{figure*}[htpb]
	\begin{center}
		\includegraphics[width=0.49\textwidth]{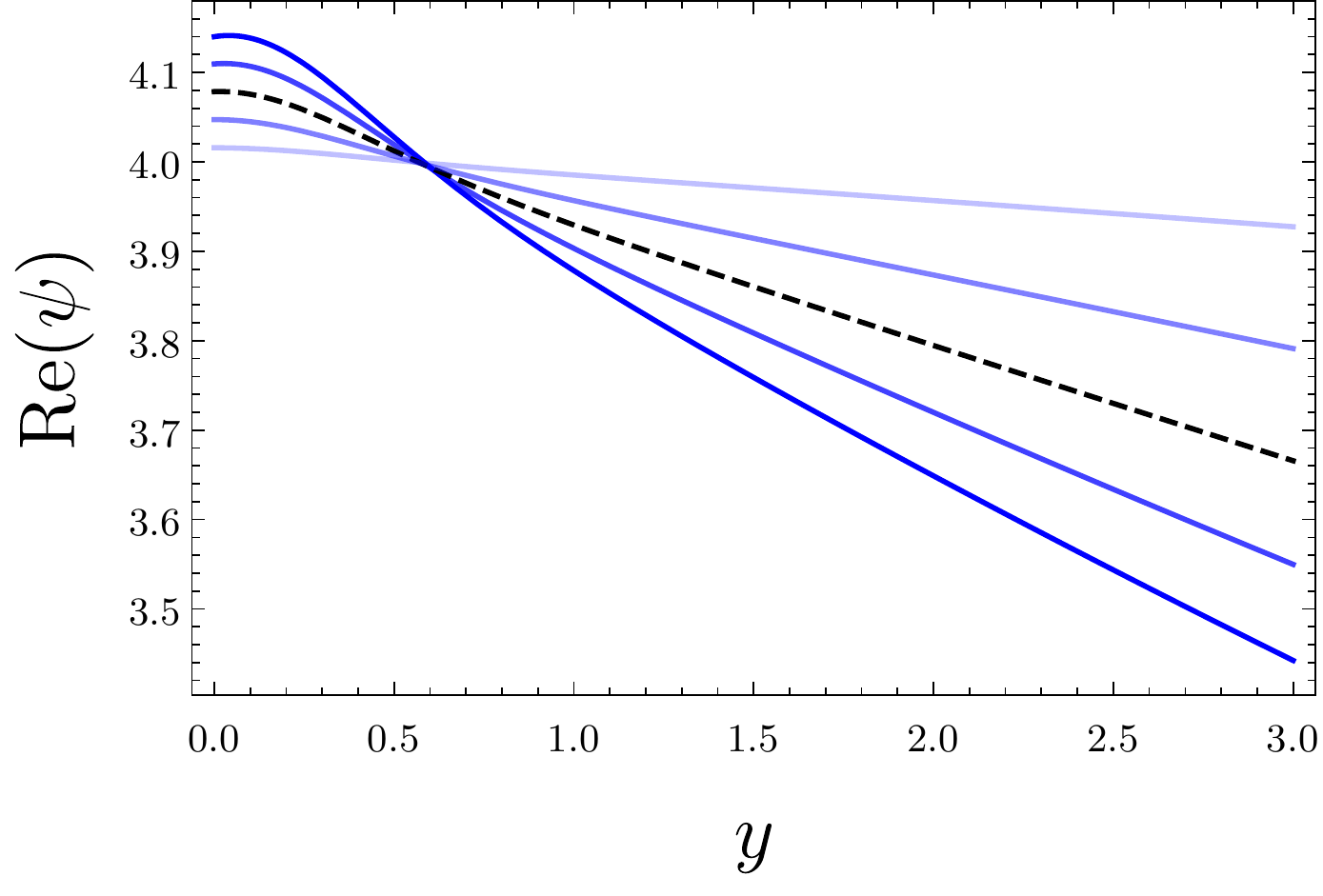}
		\includegraphics[width=0.49\textwidth]{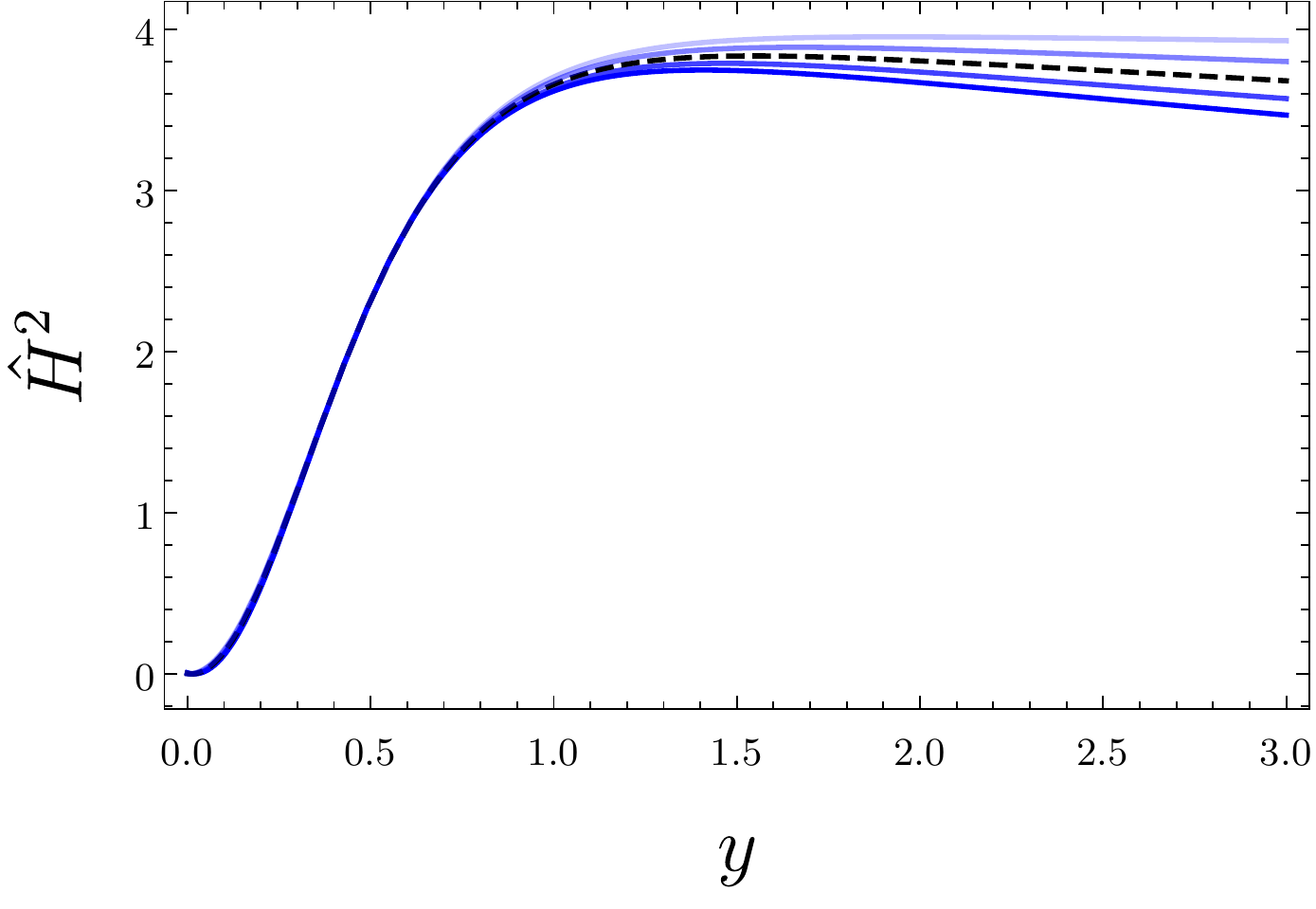}
		\includegraphics[width=0.49\textwidth]{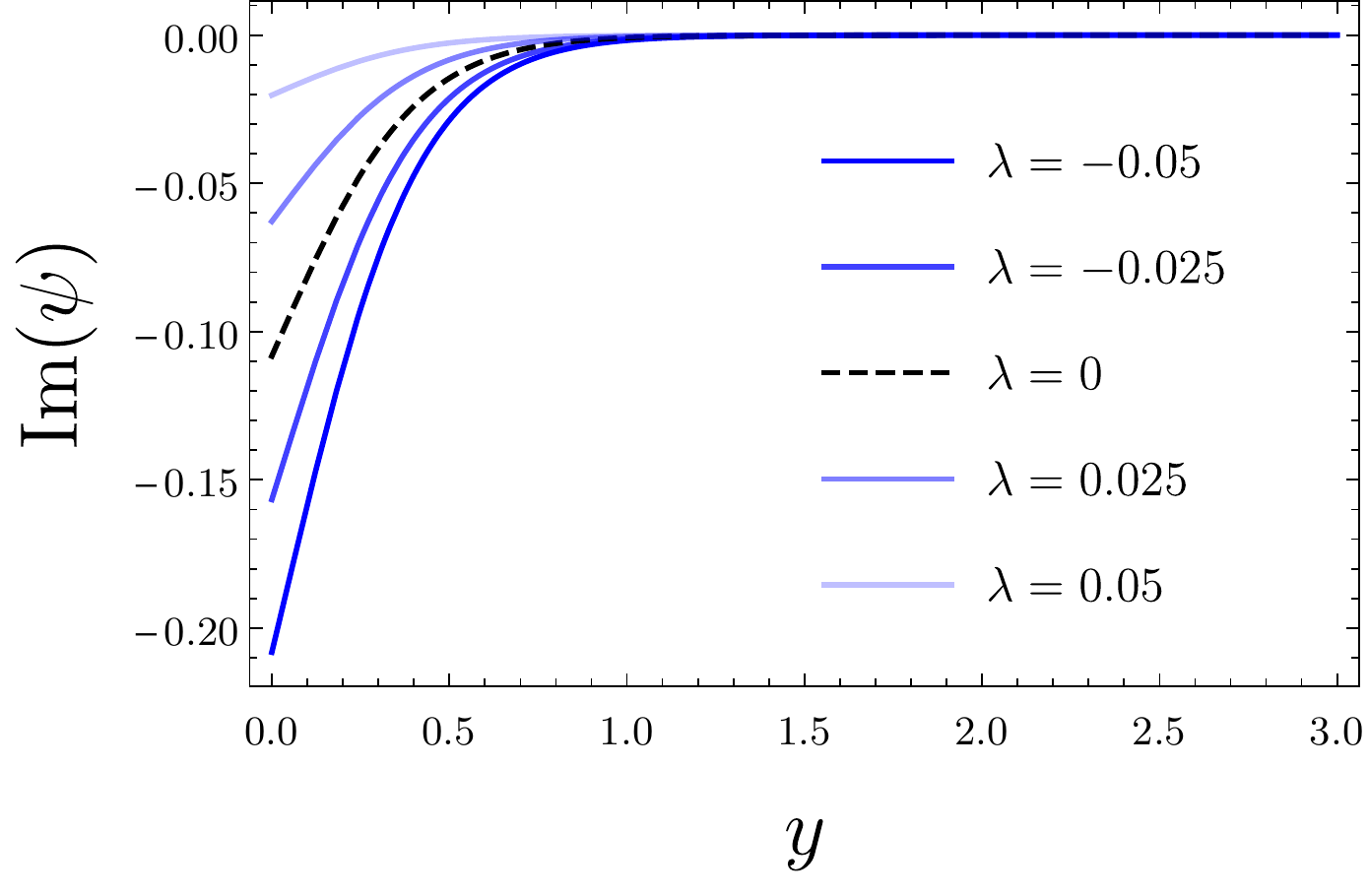}
		\includegraphics[width=0.49\textwidth]{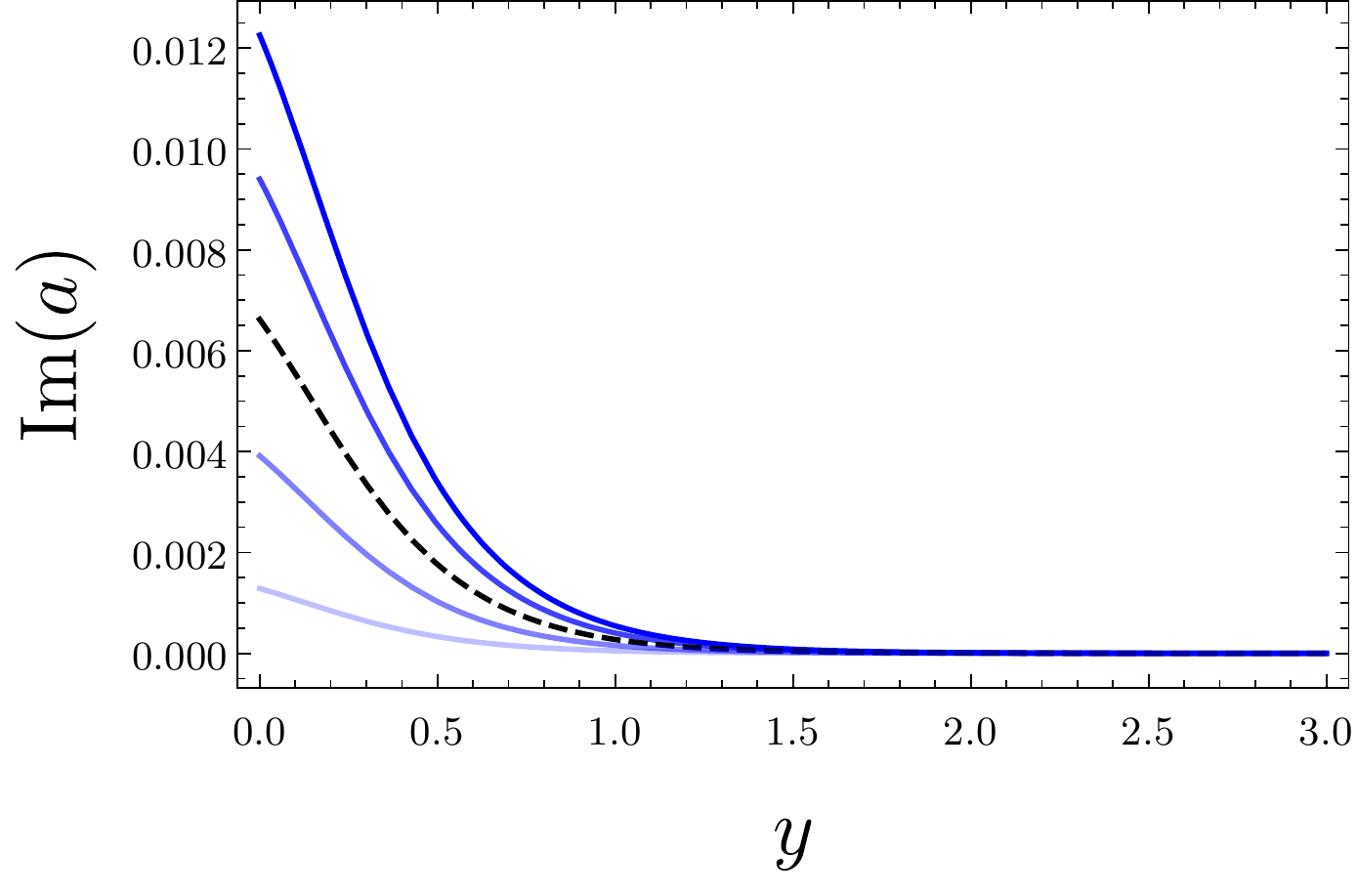}
		\caption{The evolution of complex no-boundary saddle points that describe the creation of real Lorentzian inflationary cosmologies, along the line $\tau=x_0+i y$ for five different values of $\lambda$ in the cubic theory. From left to right and from top to bottom: real part of $\psi$, Lorentzian Hubble parameter $\hat H=-\text{Re}(\dot a)/\text{Re}(a)$, imaginary part of $\psi$ and imaginary part of $a$. These curves correspond to the initial value $\ell^2|\psi(0)|=4$ in each theory and we have set $\alpha=10$. The phase of $\psi(0)$ and the turning point $x_0$ of the contour have been tuned so that the imaginary parts of $a$ and $\psi$ rapidly decay to zero, yielding a real Lorentzian solution.}
		\label{fig:HH}
	\end{center}
\end{figure*}

To find fuzzy instantons specifying the semiclassical no-boundary wave function, we follow \cite{Hartle:2008ng}
and start with a complex ansatz of the following form,
\begin{equation}
ds^2=dt^2 N(t)^2+a^2(t)d\Omega_{(3)}^2\, ,\quad \psi=\psi(t)\, .
\end{equation}
where $d\Omega_{(3)}^2$ is the metric of the 3-sphere, and $N^2$, $a^2$ and $\psi$ are complex functions of $t$. We can define a new time variable $d\tau=N dt$, so that $\tau$ takes complex values. In terms of $\tau$, the first FL equation and the equation for $\psi$ read
\begin{align}
\label{eq:tt2}
F\left(\frac{1-\dot{a}^2}{a^2}\right)&=2\alpha\ell^2\left[\psi\left(\psi-\frac{1-\dot{a}^2}{a^2}\right)+\frac{\dot a\dot\psi}{a}\right]\, ,\\
\label{eq:psi2}
\frac{1-\dot{a}^2-a\ddot{a}}{a^2}&=2\psi\, ,
\end{align}
where $\dot a=\frac{da}{d\tau}=\frac{1}{N}\frac{da}{dt}$, etc. In order to solve these equations, we search for solutions that have a regular Euclidean section near $a=0$. This happens if and only if 
\begin{equation}\label{eq:nbpa}
a(0)=0\, , \quad \dot{a}(0)=1\, ,
\end{equation}
in which case $\tau=0$ is just a smooth cap of the geometry, usually called the south pole. Thus, the only remaining boundary condition is the value of $\psi$ at the south pole, which is in general a complex number,
\begin{equation}\label{eq:nbpp}
\psi(0)=\psi_0 e^{i\theta}\, ,\quad \psi_0=|\psi(0)|\, .
\end{equation}
Now, $\psi(0)$ must be chosen in a precise way so that there exists a section of the complex solution that tends to a real Lorentzian metric. Otherwise the saddle does not contribute to the no-boundary wave function in a regime where the latter describes classical spacetime. To find those values of $\psi(0)$, first note that the solutions satisfying \req{eq:nbpa} can be analytically continued to all the complex $\tau$-plane, since one can solve \req{eq:tt2} and \req{eq:psi2} using a power series. We focus on a contour $\mathcal{C}(x_0)$ in the complex $\tau$-plane defined as follows: 
\begin{equation}
\mathcal{C}(x_0)=\{x\,|\, 0\le x\le x_0\}\cup \{x_0+i y\,|\, y\ge 0 \}\, .
\end{equation}
Numerically solving \req{eq:tt2}-\req{eq:psi2} along this contour with the boundary conditions \req{eq:nbpa}-\req{eq:nbpp} and demanding
\begin{equation}
\lim_{y\rightarrow\infty} \text{Im}\big[a(x_0+i y)\big]=0\, ,\quad \lim_{y\rightarrow\infty} \text{Im}\big[\psi(x_0+i y)\big]=0\, .
\end{equation}
we obtain an asymptotically real Lorentzian solution. Note that these are two ``classicality'' conditions, which fix two constants. It turns out that once the norm $\psi_0$ is specified, there is a unique phase $\theta$ and turning point $x_0$ along which the solution gives rise to classical behaviour. For increasing $y$ the corresponding solution soon tends to one of the inflationary solutions we considered above, with an initial value of inflation close to $\psi_0$.\footnote{A caveat in our analysis is that we considered flat FLRW metrics in the previous sections whereas the Hartle-Hawking wave function selects closed universes. However, since the scale factor grows exponentially during inflation, the effect of the curvature becomes negligible soon after the classicalization of the background.} Varying $\psi_0$ yields a one-parameter family of inflationary solutions which differ in the number of e-folds. The initial conditions specified by the no-boundary proposal thus select a line in the inflationary band in the phase space diagrams shown above. 

As an illustration, consider the case \begin{equation}
F(x)=x-\lambda x^3\, .
\end{equation}

Numerically solving the complex FL equations for several values of $\lambda$ leads to the plots shown in Fig.~\ref{fig:HH}, where we can see both the real and imaginary parts of $\psi$, the imaginary part of $a$ and the Lorentzian Hubble parameter $\hat H=-\text{Re}(\dot a)/\text{Re}(a)$. The phase of $\psi(0)$ and the turning point $x_0$ of the contour have been tuned so that the imaginary parts of $a$ and $\psi$ rapidly decay to zero, yielding an asymptotically real, Lorentzian solution. From these graphs we extract two main conclusions. 
\begin{enumerate}
\item Once the solution behaves classically --- that is, when $\text{Im}(a)<<\text{Re}(a)$ and $\text{Im}(\psi)<<\text{Re}(\psi)$ --- the values of $H$ and $\psi$ lie approximately on the slow roll line $\psi\approx H^2$. Thus the origin of classical spacetime and inflation go hand in hand \cite{Hartle:2008ng}. 
\item The values of $H$ and $\psi$ in the classical domain are almost independent of the higher-curvature corrections. We get approximately $H^2\approx \psi\approx |\psi(0)|$. 
 \end{enumerate}
 Regarding the second point, Fig.~\ref{fig:HH} shows that all curves for $\text{Re}(\psi)$ agree at $y\approx 0.6$, which is roughly when $\text{Im}(\psi)$ has decayed away. For $\hat H$, the different curves cannot be distinguished until $\text{Im}(a)<<1$.  We have verified that the same conclusions hold for larger values of the coupling $\lambda$ and $\psi_0$ and different choices of $F(x)$.  Thus,  the higher-order corrections do not seem to modify in a substantial way the initial conditions predicted by the no-boundary proposal. 
 In practice one may assume that the initial state is given by $H_{\rm in}^2\approx \psi_{\rm in}\approx \psi_0$.  
 
Not all solutions along this line are equally probable in the no-boundary state. Their relative probabilities are given by the real part of the Euclidean action which tends to a constant when the solution classicalizes \cite{Hartle:2007gi,Hartle:2008ng}. It is well known that the no-boundary wave function an sich favors a low amount of inflation. Quantum cosmological probabilities for observables such as the CMB fluctuations, however, are conditioned on the observational situation and dominated by histories with an extended phase of inflation \cite{Hawking:2006ur,Hartle:2010dq,Hertog:2013mra}.

\section{Constraints from holography}\label{sec:holoconst}

We now put forward a novel proposal to theoretically constrain the range of higher-curvature corrections to inflationary models based on the application of AdS/CFT. The idea is to add a negative cosmological constant $\Lambda$ to the action \req{Spsi} and to study the properties of the resulting theories in an AdS background. Using the holographic dictionary one can compute various quantities in the dual CFT, such as correlators of the boundary stress-energy tensor $T_{ab}$. These must satisfy certain well-known constraints in order for the dual theory to be unitary. It seems reasonable to restrict the parameter range of higher-curvature bulk theories to the range that corresponds to well-defined duals. Indeed in the AdS context, unitarity of the dual CFT must be closely related to having a bulk theory with physically reasonable properties, such as positivity of energy and causality.  In this section we derive two such `AdS/CFT constraints' on the higher-order couplings $\lambda_3$ and $\lambda_4$, resp. from the computation of the two - and three-point functions of the stress-energy tensor in the dual.   Since these are plausibly independent of the value of $\Lambda$, we then `continue' these to the inflationary theories \req{Spsi} of interest. In fact, these constraints only depend on the background curvature, which is related to the cosmological constant only on-shell.  As support for the validity of our method, we present an independent derivation directly in a de Sitter context of the first constraint stemming from the computation of the two-point function. 

We first consider the two-point function of the dual stress-energy tensor. In a CFT$_3$ this is given by 
\begin{equation}
\langle T_{ab}(x)T_{cd}(0)\rangle_{\mathbb{R}^3}=\frac{C_{T}}{|x|^6}\mathcal{I}_{ac,bd}(x)\, ,
\end{equation}
where $\mathcal{I}_{ac,bd}$ is a fixed tensorial structure and the only theory-dependent quantity is the central charge $C_T$. Unitarity implies $C_T>0$. 
Luckily, the value of this central charge for the holographic dual of the higher-curvature theories in \req{Spsi} for $\psi=0$ has already been computed \cite{ECGholo,Bueno:2018yzo,Bueno:2020odt}. That result can be easily extended to the case of $\psi\neq0$ by noticing that, whenever $\psi$ is kept fixed, it can be reabsorbed by redefining the cosmological and Newton's constants. The result then reads
\begin{equation}
C_T=C_T^{E}\left(F'(-\chi)+2\alpha\ell^2\psi\right)\, ,
\end{equation}
where $C_T^{E}=\frac{3}{\chi G\pi^3}$ is the corresponding result in Einstein gravity in AdS and $F$ is the function defined in \req{eq:Function}.
Here $\chi$ is a constant representing the curvature of the background,
\begin{equation}
R_{\mu\nu\rho\sigma}=-2\chi g_{\mu[\rho}g_{\sigma]\nu}\, ,
\end{equation}
and thus ultimately determined by the cosmological constant and the other parameters of the theory. In the case of AdS, $\chi>0$ and the positivity of the central charge implies
\begin{equation}\label{Ctconst}
\frac{C_T}{C_T^{E}}=F'(-\chi)+2\alpha\ell^2\psi>0\, .
\end{equation}

Now, since this expression only involves the evaluation of a quantity on a constant curvature background, we might as well evaluate it on a de Sitter space, for which $\chi=-H^2$. Of course, in doing so we are implicitly continuing the  cosmological constant from a negative to a positive value. In addition, de Sitter solutions in the theory \req{Spsi} have $\psi=H^2$, and therefore we arrive to the following constraint
\begin{equation}\label{eq:Const1}
F'(H^2)+2\alpha\ell^2H^2>0\, .
\end{equation}
Once again, we stress this condition has been obtained by continuing the constraint \req{Ctconst} from negative to positive curvature. This means we lose contact with the original dual theory although it should be noted that at least in some versions, dS/CFT and Euclidean AdS/CFT share the same dual \cite{Hertog:2011ky}. Further evidence for the validity of this procedure can be obtained by studying metric perturbations around de Sitter backgrounds in the theory \req{Spsi}. In that case, one finds that the linearized equations coincide with those of Einstein gravity except that $G$ is replaced by an effective Newton's constant $G_{\rm eff}$ that reads \cite{Aspects}
\begin{equation}
G_{\rm eff}=\frac{G}{F'(H^2)+2\alpha\ell^2H^2}\, .
\end{equation}
Therefore, \req{eq:Const1} must hold indeed in order to avoid $G_{\rm eff}<0$ which would imply the existence of graviton ghosts. Thus we see that the constraint \req{eq:Const1} derived from AdS holography translates into an entirely reasonable physical requirement on the theories \req{Spsi}. 

Motivated by this, we move on to the less trivial case of the 3-point function $\langle TTT\rangle$. In a 3-dimensional CFT this correlator is characterized by $C_T$ and by an additional dimensionless parameter, customarily denoted as $t_4$. Equivalently, the 3-point function can be obtained by measuring energy fluxes at the boundary of AdS after a local perturbation created by an insertion of the stress-energy tensor of the form $\epsilon_{ab}T^{ab}$ \cite{Hofman:2008ar}. The energy flux at infinity in the direction of the unit vector $\vec{n}$ is then
\begin{equation}
\langle\E(\vec{n})\rangle=\frac{E}{2\pi}\left[1+t_4\left(\frac{\left|\epsilon^{*}_{ab}n^an^b\right|^2}{\epsilon^{*}_{ab}\epsilon^{*}_{ab}}-\frac{1}{4}\right)\right]\, ,
\end{equation}
where $E$ is an overall constant. To avoid negative energy fluxes, $t_4$ must be bounded by
\begin{equation}\label{eq:t4bound}
-4\le t_4\le 4\, .
\end{equation}
Luckily, $t_4$ is also known for the holographic dual of \req{Spsi} for $\psi=0$ \cite{ECGholo,Bueno:2020odt}. Again, an extension to the case of $\psi\neq 0$ yields
\begin{equation}
t_4=-\frac{210\chi F''(-\chi)}{F'(-\chi)+2\alpha\ell^2\psi}\, .
\end{equation}
As before, this only depends on the background curvature, and hence replacing $\chi=-H^2$ and $\psi=H^2$, and taking into account \req{eq:t4bound}, we arrive at the following constraint 
\begin{equation}\label{eq:Const2}
-4\le \frac{210H^2 F''(H^2)}{F'(H^2)+2\alpha\ell^2H^2}\le 4\, .
\end{equation}
This is a much stronger bound than \req{eq:Const1}. It is clearly violated in theories where $F'(H^2)+2\alpha\ell^2H^2$ approaches zero somewhere.
 
 It would be very interesting to better understand the physical implications in the bulk of the violation of \req{eq:Const2}. In the case of Lovelock gravities in AdS, it is known that the unitarity constraints arising from the positivity of energy fluxes coincide with causality constraints in the bulk \cite{Brigante:2007nu,deBoer:2009pn,Camanho:2009vw,Buchel:2009sk,Camanho:2009hu}. Hence \req{eq:Const2} in de Sitter space might similarly be related to causality. It was noticed in Refs.~\cite{Hofman:2009ug,Myers:2010jv}, however, that in theories other than Lovelock gravity, this correspondence does not seem to hold. This is because the terms with higher-order derivatives dominate at large momentum, yielding a dispersion relation without corrections. We briefly discuss this in the next section when we compute gravitational perturbations. Clearly the physical meaning in the bulk of \req{eq:Const2} deserves further study. For now, we investigate the consequences of these two bounds upon the assumption that they must hold in order for the theory to be physically reasonable.

To be explicit we consider the relevant case of a single cubic term in the action, corresponding to
\begin{equation}\label{FHex}
F(H^2)=H^2-\lambda_3 \ell^4 H^6\, .
\end{equation}
The first constraint \req{eq:Const1} implies
\begin{equation}\label{Constex}
1+2\alpha (\ell H)^2-3\lambda_3 (\ell H)^4>0\, .
\end{equation}
Now, strictly speaking, this is a condition that applies in de Sitter vacua. That is, in this final derivation we have assumed a positive cosmological constant on top of the action \req{Spsi}. Nevertheless, it is easy to see that the same condition should hold throughout the quasi-de Sitter phases of interest. In particular, if $1+2\alpha (\ell H)^2-3\lambda_3 (\ell H)^4<0$ during inflation, with $\epsilon<<1$, one necessarily runs into ghostly gravitational waves. Note that in the model given by \req{FHex}, with $\lambda_3>0$, we have $H^2\le\frac{1}{\ell^2\sqrt{\lambda_3}}$ during inflation, with the upper bound corresponding to the exact (unstable) de Sitter solution. Hence \req{Constex} is satisfied for an arbitrary number of e-folds as long as $\lambda_3\le \alpha^2$.\footnote{The bound is approximately the same if one only demands \req{Constex} holds for $N\le 60$.} Notice that $\lambda_3\lesssim \alpha^2$ means that the higher-order terms can dominate when $N$ is large enough. On the other hand, if $\lambda_3<0$ there is no constraint from \req{Constex}, although in that case there is a lower bound on $\lambda_{3}$ in order to achieve a sufficient number of e-folds according to \req{eq:F1max}.

The conditions in \req{eq:Const2} in turn read 
\begin{align}
1+2\alpha (\ell H)^2-318 \lambda_3 (\ell H)^4\ge &0\, ,\\
1+2\alpha (\ell H)^2+312 \lambda_3 (\ell H)^4\ge &0\, ,
\end{align}
These are much stronger than \req{Constex}. During the slow-roll phase we can obtain $H(N)$ by using \req{eqn:F5psi} and $H^2\approx \psi$. Imposing that the conditions above hold for $N\le 60$, leads to the constraints on $\lambda_3$ shown in Table~\ref{tab1}. These strongly restrict the range of allowed values of $\lambda_{3}/\alpha^2$, eliminating in particular the possibility of large corrections to $R^2$ inflation. Table \ref{tab1} also shows the bounds for other choices of $F(H^2)$ with the same cubic term but differing in the asymptotic behaviour for $H\rightarrow\infty$. As we see, the constraints arising from \req{eq:Const1} are similar in order of magnitude in all cases, and roughly imply $|\lambda_{3}|/\alpha^2\lesssim 10^{-4}$. However, the quantity that determines whether higher derivative corrections are relevant during inflation is $N^2|\lambda_{3}|/\alpha^2$ and this can approach $1$ at the beginning of inflation $N\sim 60$. Therefore, the possibility of $\mathcal{O}(1)$ corrections is not ruled out. 

Finally, we also show in Table \ref{tab1} the bounds corresponding to a quartic correction, $F(H^2)=H^2+\lambda_{4}\ell^6 H^8$. In this case, the effect of the corrections is of the order of $N^3|\lambda_{4}|/\alpha^3$, and we again conclude that this quantity can be of order $\mathcal{O}(1)$ at the beginning of inflation. Of course one may similarly consider the effect of a combination of higher-order terms (\textit{e.g.} cubic and quartic together), in which case the conditions \req{eq:Const2} constrain the couplings $\lambda_{n}$ to a certain region of the parameter space --- see Fig.~\ref{fig:constraintspace} below. Finally we note that imposing the constraints \req{eq:Const1} and \req{eq:Const2} to hold for $N\le 60$ is the weakest condition in order to keep things under control in the observationally relevant range. However, if inflation lasts for more than 60 e-folds then one may want to demand that \req{eq:Const1} and \req{eq:Const2} hold for higher $N$, in which case the constraints on the couplings would be even stronger.

\begin{table}[t!] 
\def\arraystretch{1.8}
\centering
	\begin{tabular}{|c|c|c|}
		\hline
	$\displaystyle F(H^2)$	&    Constr. \req{eq:Const1}  &  Constr. \req{eq:Const2}  \\ 
		\hline\hline 
	$\displaystyle H^2-\lambda_3 \ell^4 H^6$	 & $\displaystyle \frac{\lambda_3}{\alpha^2}<1$ & $\displaystyle-1.5\le 10^4\frac{\lambda_3}{\alpha^2}\le 1.7$\\  [6pt]
		 \hline
	$\displaystyle H^2\left(1-2\lambda_3 \ell^4 H^4\right)^{1/2}$	& $\displaystyle \frac{\lambda_3}{\alpha^2}<0.00074$ & $\displaystyle-2.4\le10^4\frac{\lambda_3}{\alpha^2}\le 1.2$\\  [6pt]	 
	\hline
	$\displaystyle H^2\left(1+2\lambda_3 \ell^4 H^4\right)^{-1/2}$	& $- $ & $\displaystyle-0.72\le10^4\frac{\lambda_3}{\alpha^2}\le 10$\\  [6pt]
	\hline
	$\displaystyle H^2+\lambda_4 \ell^6 H^8$	& $\displaystyle \frac{\lambda_4}{\alpha^3}>-\frac{8}{27}$ & $\displaystyle-2.1\le10^6\frac{\lambda_4}{\alpha^3}\le 1.9$\\  [6pt]
	\hline
	\end{tabular}
	\caption{Constraints to the higher-order couplings $\lambda_3$ and $\lambda_{4}$ coming from the unitarity conditions \req{eq:Const1} and \req{eq:Const2}. In each case we demand that the corresponding condition is satisfied during inflation for at least $N\le 60$. }
	\label{tab1}
\end{table}

\section{Perturbations}\label{sec:pert}

With the background analysis, the initial conditions, and (some of) the theoretical constraints in place, we finally turn to the phenomenology of the higher-curvature extensions \req{Spsi} of $R^2$ inflation. In this section we compute the linear scalar and tensor perturbations in some of the theories \req{Spsi}, around the inflationary backgrounds discussed in Section~\ref{sec:slowroll}. We focus on the phenomenologically relevant cubic and quartic terms given in resp. \req{eq:cECG} and \req{eq:quartic}. We calculate the corrections to the power spectrum of tensor and scalar perturbations at leading order in the parameters $\lambda_3$ and $\lambda_4$. This is a reasonable choice since the theoretical analysis carried out in the previous section as well as current observational constraints imply that the hypothetical higher-curvature corrections must be small. One remarkable property of the densities \req{eq:cECG} and \req{eq:quartic} is that their linearized equations around FLRW backgrounds are of second order in time derivatives,  greatly simplifying the analysis. 
In what follows we perform in detail the analysis of perturbations in the cubic theory, while the computations for the quartic theory can be found in the Appendix \ref{app:pert}. 

\subsection{Tensor perturbations}

Gravitational waves are easier to study since they do not couple to the scalar field $\psi$. They can be written in terms of the following metric perturbation,

\begin{equation}
ds^2=a(\tau)^2\left(-d\tau^2+\left(\delta_{ij}+h_{ij}\right)dx^{i}dx^{j}\right)\, ,
\end{equation}
where we are using the conformal time $\tau$. Additional gauge freedom allows us to choose $h_{ij}$ to be transverse and traceless:
\begin{equation}
\tensor{h}{^{i}_{i}}=0, \quad \partial_{i}\tensor{h}{^{i}_{j}}=0\, .
\end{equation}

Working in momentum space and separating $h_{ij}$ in its two polarizations, we can write
\begin{equation}
h_{ij}=\int d^3k h^{(+)}_ke^{(+)}_{ij}e^{i \vec{k}\cdot \vec{x}}+\int d^3k h^{(\times)}_ke^{(\times)}_{ij}e^{i \vec{k}\cdot \vec{x}}\, ,
\end{equation}
where for both polarization tensors

\begin{equation}
\tensor{e}{^{(\lambda)i}_{i}}=0, \quad k^{i}e^{(\lambda)}_{ij}=0\, .
\end{equation}
In addition, they are orthogonal and normalized to unity:

\begin{equation}
\left(e^{(\lambda)}_{ij}\right)^*e^{(\lambda')}_{ij}=\delta_{\lambda\lambda'}\, .
\end{equation}

Expanding the action quadratically in the perturbation and integrating by parts several times, we are left with the following action for each of the two polarizations (we omit the $(+)$ and $(\times)$ labels as well as the integration over $k$ from now on),

\begin{equation}
\label{Sh}
S_{h}=\frac{1}{32\pi G}\int d\tau\frac{a^2}{2}\left[f  h_{k}'^2-\omega^2 h_k^2\right]\, ,
\end{equation}
where $f$ and $\omega^2$ are the following functions, 

\begin{widetext}
\begin{align}\notag
\omega^2=&k^2 \left(1+2 \alpha  \ell^2 \psi -3 \left(38 H^4-70 H^2
   \psi +33 \psi ^2\right) \ell^4 \lambda _3-\frac{9 \ell^2 \lambda _3}{2 \alpha } \left(-5 H^2+4 \psi +\left(13 H^6-12 H^4 \psi
   \right) \ell^4 \lambda _3\right)\right)\\\label{eq:omega}
   &+\frac{6 k^4}{a^2} \left(\psi-H^2 \right) \ell^4\lambda _3\, ,\\
 f=&1+2 \alpha \ell^2  \psi+\left(42 H^4+27 \psi ^2-\frac{9
   H^2 (1+16 \alpha  \psi )}{2 \alpha }\right)\ell^4 \lambda _3+\frac{9\ell^6H^6 \lambda _3^2}{2 \alpha } -\frac{6 k^2}{a^2} \left(H^2-\psi \right)\ell^4 \lambda _3\, .\label{eq:ftensor}
\end{align}
\end{widetext}
Notice that $h_k$ does not have a canonically normalized kinetic term, but we can achieve that by performing the redefinition
\begin{equation}\label{hkchik}
h_k=\frac{2\chi_k}{M_{\rm P} \beta}\, ,\,\, \text{where}\,\,\, \beta=a\sqrt{f}\, ,
\end{equation}
and where $M_{\rm P}=(8\pi G)^{-1/2}$ is the reduced Planck mass. In this case, performing again an integration by parts in \req{Sh}, we get the following action for $\chi_k$:

\begin{equation}
\label{Schi}
S_{h}=\int d\tau\frac{1}{2}\left[\chi_{k}'^2-\hat\omega^2 \chi_k^2\right]\, ,\quad  \hat\omega^2=\omega^2/f-\frac{\beta''}{\beta}\, .
\end{equation}
Then, $\chi_k$ satisfies the equation

\begin{equation}\label{eq:chik}
\chi_k''+ \hat\omega^2 \chi_k=0\, .
\end{equation}
The last step is to evaluate the quantity $ \hat\omega^2$ on the slow roll phase. We will perform an expansion in the slow roll parameter so that we will only keep the leading terms in $\epsilon$. Likewise, we will work perturbatively in $\lambda_{3}$, which means that we perform an expansion in the parameter
\begin{equation}
\zeta=\frac{\lambda_{3}}{\alpha^2\epsilon^2}\, ,
\end{equation}
and we stay at linear order in $\zeta$. The computation of $\omega^2-\frac{\beta''}{\beta}$ presents an added challenge in the fact that the terms linear in $\epsilon$ terms vanish --- as is known to happen for $R^2$ inflation --- so that we need to go to second order in the slow roll expansion.  
This can be done by introducing additional slow roll parameters that capture higher derivatives of $H$. There are many ways of doing this, but we have checked that they yield the same result. For instance, we may define 
\begin{equation}
\epsilon_{n}=(-1)^n\frac{H^{(n)}}{H^{n+1}}\, ,
\end{equation}
where the derivatives are taken with respect to the cosmic time $t$. Then we can achieve higher-order slow roll approximations by assuming that $\epsilon_n\approx 0$ for some $n$. In particular, the first-order approximation can be obtained from $\epsilon_2=0$.  The second-order approximation can be obtained from $\epsilon_3=0$, which yields the equation

\begin{equation}
6+4 \epsilon -3 \alpha  \epsilon  (6+5 \epsilon ) \ell^2 H^2-6 \zeta \alpha ^2 \epsilon ^2 (1+2 \epsilon )
  \ell^4 H^4=0\, .
\end{equation}
From here we get

\begin{equation}\label{Heps2}
\ell^2 H^2=\frac{1-\zeta/9}{3\alpha\epsilon}-\frac{9+5\zeta}{162\alpha}+\mathcal{O}(\epsilon,\zeta^2)\, .
\end{equation}
On the other hand, we also need to approximate the value of $aH$ at order $\epsilon^2$. 
Now $aH$ satisfies the following equation
\begin{equation}
(aH)'=(aH)^2(1-\epsilon)\, .
\end{equation}
Assume further that this equation has the following solution when $\epsilon<<1$,
\begin{equation}\label{eq:aHsol}
\frac{1}{aH}=-\tau\left(1-\epsilon+c\epsilon^2\right)\, ,
\end{equation}
for some constant $c$. Taking into account that $\epsilon'\sim\mathcal{O}(\epsilon^2)$, the equation above then implies that 
\begin{equation}
c\epsilon^2+\frac{\epsilon'}{aH}=\mathcal{O}(\epsilon^3)\, .
\end{equation}
Thus, this constant must be given by
\begin{equation}
c=-\lim_{\epsilon\rightarrow0}\frac{\epsilon'}{\epsilon^2aH}\, .
\end{equation}
Using \req{slowroll2} we can compute the derivative of $\epsilon$ and we get
\begin{equation}
c=-4+\lim_{\epsilon\rightarrow 0}\frac{2F'(H^2)}{3\alpha \ell^2 H^2\epsilon}\, .
\end{equation}
Finally, taking into account \req{Heps2} we obtain 
\begin{equation}
c=-2-\frac{4\zeta}{9}\, .
\end{equation}
Thus, we have derived $aH$ and $H$ at second order in $\epsilon$ and at first order in $\zeta$. We can now insert these expressions in \req{eq:chik} and we obtain the following result after expanding once again:

\begin{equation}\label{eq:chik2}
\chi_k''+\left(\tilde k^2-\frac{1}{\tau^2}\left(\nu^2-\frac{1}{4}\right)\right) \chi_k=0\, ,
\end{equation}
where
\begin{equation}
\tilde k^2=k^2\left(1+\frac{3\zeta\epsilon^2}{2}\right)\, ,\quad \nu^2=\frac{9}{4}+3(3+\zeta)\epsilon^2\, .
\end{equation}

The equation \eqref{eq:chik2} is solved in terms of the Hankel functions $H^{(1,2)}_{\nu}(x)$ as follows,
\begin{equation}\label{eq:Hankelsol}
\chi_k=\sqrt{-\tau}\left[c_1 H_{\nu}^{(1)}(-\tilde k \tau)+c_2 H_{\nu}^{(2)}(-\tilde k \tau)\right]\, .
\end{equation}
Now the integration constants must be chosen so that $\chi_k$ becomes a canonically-normalized positive-frequency plane wave on sub-Hubble scales $k>>aH$ $(-k\tau>>1)$:
\begin{equation}\label{eq:chiklarge}
\chi_k\approx\frac{e^{-i \tilde{k}\tau}}{\sqrt{2\tilde{k}}} \quad \text{when} \quad k>>aH\, .
\end{equation}
According to this formula, the gravitational waves at large momentum $k$ would have a frequency $\omega^2=k^2\left(1+\frac{3\zeta\epsilon^2}{2}\right)$, so one would need to take $\lambda_{3}<0$ in order to avoid superluminal propagation. However, one has to be cautious when taking the limit $k\rightarrow\infty$. 
In fact, the equation \req{eq:chik2} does not capture the limit of large momentum correctly. As we can see in Eqs.~\req{eq:omega} and \req{eq:ftensor} there are terms which are proportional to $\lambda_{3}$ that contain higher powers of $k$. As it turns out, these terms do not appear in \req{eq:chik2} because they are subleading in the expansion in $\zeta$ and $\epsilon$. However, they will become relevant if $k$ is large enough. Indeed in order to neglect those terms we must have
\begin{equation}
\frac{k^2}{(aH)^2}\zeta\epsilon^2<<1\, .
\end{equation}
Note that even if $\zeta\sim 1$ it is still possible to satisfy both $k>>aH$ and the constraint above. Nevertheless, in the limit $k\rightarrow\infty$ those terms become dominant and have to be taken in account. We may compute then the velocity of the wave as 
\begin{equation}
v^2=\lim_{k\rightarrow\infty}\frac{\hat\omega^2}{k^2}\, ,
\end{equation}
where $\hat\omega^2$ was defined in \req{Schi}. If we perform this limit without any approximation we find the result $v=1$, so that gravitational waves move at the speed of light, contrarily as we found before. This would also imply that, at least in this setup, no causality violations are found. This is quite similar to the results in \cite{Hofman:2009ug,Myers:2010jv} in the case of AdS. Nevertheless, it would be interesting to explore if causality bounds can be obtained by analyzing perturbations on less symmetric asymptotically dS solutions. 

Now, let us continue with the discussion of the solution of the equation \req{eq:chik2}. In the range $(aH)/(\sqrt{\zeta}\epsilon)>>k>>aH$ the solution \req{eq:Hankelsol} tends to the plane wave \req{eq:chiklarge} if the integration constants are chosen as $c_1=\frac{\sqrt{\pi}}{2}e^{i(\nu+1/2)\pi/2}$, $c_2=0$.
This leads to

\begin{equation}
\chi_k=\sqrt{-\tau}\frac{\sqrt{\pi}}{2}e^{i(\nu+1/2)\pi/2}H_{\nu}^{(1)}(-\tilde k \tau)\, .
\end{equation}

On the other hand, on super-Hubble scales $k<<aH$ the wave freezes and we have 

\begin{equation}
\chi_k\approx e^{i(\nu-1/2)\pi/2}2^{\nu-3/2}\frac{\Gamma(\nu)}{\Gamma(3/2)}\frac{(-\tilde k\tau)^{1/2-\nu}}{\sqrt{2\tilde k}}\, .
\end{equation}
Now, using \eqref{hkchik} we recover the metric perturbation $h_k$ in the super-Hubble regime. Taking into account that $\nu=3/2+\mathcal{O}(\epsilon^2)$, we have 

\begin{equation}\label{eqn:h}
|h_k|\approx \frac{2}{M_{\rm P}a \sqrt{2fk}}\left(\frac{k}{aH}\right)^{1/2-\nu} \quad \text{when} \quad k <<aH\, ,
\end{equation}
where we are neglecting the $\mathcal{O}(\epsilon^2)$ terms in the amplitude of $h_k$, but not in the exponent of $k$, $1/2-\nu$. 
It is interesting to note at this point that the function $f$ appearing in the denominator of this expression can be shown to become negative during inflation if the bound \req{eq:Const1} is not satisfied. This would lead to a catastrophic breakdown of the linearized perturbations, which shows the importance of \req{eq:Const1}.

For the power spectrum, given by
\begin{equation}\label{eqn:PT}
\mathcal{P}_{T}(k)=\frac{k^3}{2\pi^2}\left(\left|h_{k}^{(+)}\right|^2+\left|h_{k}^{(\times)}\right|^2\right)\, .
\end{equation}
we find that
\begin{equation}\label{eqn:PTdef}
\mathcal{P}_{T}(k)=A_{T}^2\left(\frac{k}{aH}\right)^{n_{T}}\, .
\end{equation}
where the  amplitude $A_T$ and spectral index $n_{T}$ of tensor perturbations read
\begin{align}
 A_{T}^2=&\frac{8}{M_{\rm P}^2f}\left(\frac{H}{2\pi}\right)^2\approx \frac{12\epsilon\left(1+\zeta/9\right)}{M_{\rm P}^2}\left(\frac{H}{2\pi}\right)^2\, ,\\
 n_{T}=&3-2\nu\approx -2(3+\zeta)\epsilon^2\, .
\end{align}
\noindent
Finally, taking into account the relation \req{slowroll2} and the evolution of $\psi$ as a function of the number of e-folds given by \req{eqn:F5psi}, we obtain, at linear order in $\lambda_3$, 
\begin{equation}
\epsilon\approx \frac{1}{2N}-\frac{4\lambda_{3}}{27\alpha^2}N\, .
\end{equation}
Thus, we obtain our final answer for the spectral index expressed in terms of the number of e-folds before the end of inflation,
\begin{equation}
n_T=-\frac{3}{2N^2}-\frac{10\lambda_{3}}{9\alpha^2}\, .
\end{equation}
Notice that for $\lambda_{3}=0$ we recover the known result for $R^2$ inflation. 

The approach presented here for the cubic action \req{eq:cECG} can be straightforwardly extended to include the quartic corrections from \req{eq:quartic}. We simply state the final result and give more details in Appendix~\ref{app:tensor}

\begin{equation}
n_T = -\frac{3}{2N^2}-\frac{10\lambda_{3}}{9\alpha^2} +\frac{2\lambda_4 N}{\alpha^3} \, .
\end{equation}

Let us note that the density $\R^{B}_{(4)}$ from \req{eq:quartic} in principle contributes to the tensor perturbations but only at higher order in the slow-roll expansion.

\subsection{Scalar perturbations} \label{sec:scalarperturbations}
We now turn to the more involved case of scalar perturbations. We assume that the inflaton $\psi$ is perturbed so that
\begin{equation}
\psi\rightarrow\psi+\delta\psi\, ,
\end{equation}
while scalar metric perturbations can always be written in a gauge such that they take the following form

\begin{equation}
ds^2=a(\tau)^2\left[-d\tau^2(1+2\Phi)+(1-2\Psi)\delta_{ij}dx^idx^j\right]\, .
\end{equation}
Our goal shall be to compute the power spectrum of the comoving curvature perturbation, given by

\begin{equation}\label{comcurv}
\mathcal{R}=\Psi+aH\frac{\delta\psi}{\psi'}\, .
\end{equation}
It is interesting to note that this quantity is not only gauge invariant, but also frame invariant under redefinitions of $g_{\mu\nu}$ and $\psi$ of the form $\tilde g_{\mu\nu}=f(\psi) g_{\mu\nu}$, $\tilde \psi=g(\psi)$. Now we decompose the perturbations in Fourier modes,
\begin{align}
\Psi=&\int d^3k\,\Psi_k e^{i\vec{k}\cdot\vec x}\, ,\\
\Phi=&\int d^3k\, \Phi_k e^{i\vec{k}\cdot\vec x}\, ,\\
\delta\psi=&\int d^3k\, \delta\psi_k e^{i\vec{k}\cdot\vec x}\, .
\end{align} 
and we evaluate the Einstein equations $\mathcal{E}_{\mu\nu}$ given by \req{eq:Einstein} and the equation of the scalar \req{eq:psieom}. For each mode $k$, we find that the linearized Einstein's equations have the following structure
\begin{equation}
\mathcal{E}_{\tau\tau}\, ,\quad \mathcal{E}_{\tau i}=k^i A\, ,\quad \mathcal{E}_{ij}=B k^{i}k^{j}+C g_{ij}\, .
\end{equation}
The explicit form of the four components $\mathcal{E}_{\tau\tau}$, $A$, $B$ and $C$ is shown in the appendix. These, together with Eq.~\req{eq:psieom}, would make a total of five equations of motion; however, the Bianchi identity of the Einstein's equations imply two relations among these equations, so that only three of them are independent.  In addition, instead of working with the equation \req{eq:psieom}, we find it more convenient to consider the equation

\begin{equation}\label{eq:scalarpert}
6\alpha \ell^2 \nabla^2\psi-12\psi+\lambda_{3} \ell^2 \tensor{\mathcal{E}}{^{(3)\mu}_{\mu}}=0\, ,
\end{equation}
which is obtained by combining \req{eq:psieom} with the trace of Einstein's equations \req{eq:Einstein}\, .

In order to solve these equations, it proves useful to perform first the following change of variables,
\begin{align}
\Psi_k= &\, \tilde\Psi_k+\frac{\alpha \ell^2 \delta\psi_k}{1+2\alpha \ell^2\psi}\, ,\\
\Phi_k= &\, \tilde\Phi_k-\frac{\alpha \ell^2 \delta\psi_k}{1+2\alpha\ell^2 \psi}\, .
\end{align} 
The reason for this redefinition is that, in the case of $\lambda_{3}=0$, $\tilde\Psi_k$ and $\tilde\Phi_k$ correspond to the metric perturbation in the Einstein frame.  Now, when $\lambda_{3}=0$ one can check that the equation $B=0$ implies that $\tilde\Phi_k=\tilde\Psi_k$, so that both variables are equal. It turns out that this property is broken when $\lambda_3\neq 0$. However, since we treat $\lambda_3$ as a small coupling we can write
\begin{equation}
\tilde\Phi_k=\tilde\Psi_k+\lambda_3 \ell^4\varphi_k\, .
\end{equation}
The form of the equations of motion is very cumbersome, but since we are only interested in the slow roll regime and small couplings, we can expand them in $\epsilon$ and in $\zeta=\lambda_{3}/(\alpha^2\epsilon^2)$, using the relations \req{Heps2} and \req{eq:aHsol}. Doing so, the equations $A=0$ and $B=0$ become, at leading order in $\epsilon$ and first order in $\zeta$,
\begin{align}
\notag
0=&\frac{4 (9-\zeta )}{27 \epsilon }\left(a H  \tilde\Psi _k+ \tilde\Psi _k'\right)+ \alpha \ell^2 \epsilon  \left(a H
  (6+\zeta ) \delta \psi _k- \zeta  \delta \psi _k'\right)\\\label{eq:A=0}
  &+\frac{4}{3} a H \alpha ^2 \ell^4 \epsilon  \zeta  \varphi_{k}\\\notag
0=&\epsilon\zeta\bigg[\frac{2}{3} \alpha ^2 \ell^4 \varphi_{k}+\alpha \ell^2 \delta \psi_k+\frac{1}{6} \left(-2+\frac{3 k^2}{(aH)^2}\right)   \tilde\Psi _k+\frac{\tilde\Psi _k'}{a H}\\\label{eq:B=0}
   &+\frac{ \tilde\Psi _k''}{2 (aH)^2}\bigg]\, .
\end{align}
In each case we are only keeping the leading term of every field. For instance, in the first equation there are subleading terms of the form $\sim aH \tilde\Psi_k$ but these are negligible in comparison with the leading term which is of order $\sim aH \tilde\Psi_k/\epsilon$. 
Now, looking at the powers of $\epsilon$ in \req{eq:A=0} it is clear that $\tilde\Psi_{k}\sim\mathcal{O}(\epsilon^2\delta \psi_k,\epsilon^2 \varphi_k)$. Then, we can neglect the terms with $\Psi_k$ in \req{eq:B=0} and we get

\begin{equation}\label{eq:varphisol}
\varphi_{k}\approx -\frac{3}{2\alpha \ell^2}\delta\psi_{k}\, .
\end{equation}
On the other hand, we are interested in the evolution of perturbations for super-Hubble scales, $k<<aH$, and in this regime we expect that the perturbations freeze so that $|\Psi_{k}'|<<aH|\Psi_{k}|$, $|\delta\psi_{k}'|<<aH|\delta\psi_{k}|$ --- a expectation that is confirmed later. Thus, neglecting the derivatives in \req{eq:A=0} and using the result \req{eq:varphisol} we obtain 
\begin{equation}\label{eq:Psiksol}
\tilde\Psi_{k}=-\frac{3}{4}(6-\zeta/3) \alpha \ell^2 \epsilon^2 \delta\psi_k\, .
\end{equation}
Now, consider equation \req{eq:scalarpert}, which at next-to-leading order in $\epsilon$ reads
\begin{widetext}
\begin{align}\notag
0=&\delta \psi _k''+2 a H
   (1+\epsilon ) \delta \psi _k'+\left(k^2-\frac{4}{3} a^2 H^2 \epsilon  \zeta \right) \delta \psi _k+\frac{(3-\epsilon ) \zeta  \tilde\Psi _k''}{9
   \alpha \ell^2}+\left(4 a H (6-\epsilon )+\frac{a H (84-248 \epsilon ) \zeta }{9 }\right) \frac{\tilde\Psi _k'}{9 \alpha \ell^2}\\
   &+\left(a^2 H^2 (4-2 \epsilon )+ \zeta\left(\frac{a^2 H^2(24-98 \epsilon )}{27}+\frac{k^4 \epsilon }{9 a^2 H^2  }-\frac{k^2 (-3+13 \epsilon)}{27 }\right) \right) \frac{\tilde\Psi _k}{\alpha \ell^2}+4 a^2 H^2 \alpha \ell^2 \epsilon ^2 \zeta  \varphi_k+\frac{2}{3} a H \alpha \ell^2 \epsilon ^2 \zeta  \varphi_k'\, . \label{eq:scalarpert2}
\end{align}
\end{widetext}
Inserting \req{eq:Psiksol} and \req{eq:varphisol} in \req{eq:scalarpert2} we realize that the terms with $\tilde\Psi_{k}$ and $\varphi_{k}$ are of order $\mathcal{O}(\epsilon^2)$, and hence, negligible at the order we are working. Thus, we obtain the master equation for $\delta\psi_k$:

\begin{equation}
\delta \psi _k''+2 a H (1+\epsilon ) \delta \psi_k'+\left(k^2-\frac{4}{3} a^2 H^2 \epsilon  \zeta \right) \delta \psi _k=0
\end{equation}
Finally, we redefine $\delta\psi_k$ in terms of a new variable,
\begin{equation}\label{dpsired}
\delta\psi_k=A_0\sqrt{1+2\alpha\ell^2 \psi}\frac{\delta\chi_k}{a}\, ,
\end{equation}
where $A_0$ is a normalization constant that will be determined later. Then, we derive the following equation for $\chi_k$:
\begin{equation}\label{eq:chik3}
\delta\chi_k''+\left(k^2-\frac{1}{\tau^2}\left(\nu_{s}^2-\frac{1}{4}\right)\right) \delta\chi_k=0\, ,
\end{equation}
where 

\begin{equation}
\nu_s^2=\frac{9}{4}+6\epsilon\left(1+\frac{2\zeta}{9}\right)\, .
\end{equation}

Following the same discussion as in the previous section, we conclude that the properly normalized solution is given by
\begin{equation}\label{eq:chiksol2}
\delta\chi_k=\sqrt{-\tau}\frac{\sqrt{\pi}}{2}e^{i(\nu_s+1/2)\pi/2}H_{\nu_{s}}^{(1)}(-k \tau)\, ,
\end{equation}
so that on super-Hubble scales we have 

\begin{equation}\label{chisupH}
|\delta\chi_k|\approx \frac{(- k\tau)^{1/2-\nu_{s}}}{\sqrt{2 k}} \, .
\end{equation}
To determine the normalization constant we have to study the solution on sub-Hubble scales and make sure that $\delta \chi_k$ has a canonically normalized kinetic term. However, due to the gauge freedom there is no well-defined Lagrangian for this variable. In particular, the redefinitions of the other variables $\Psi_k$, $\Phi_k$ will affect the kinetic term of $\delta \chi_k$. Instead, one should treat the three quantities $\Psi_{k}$, $\Phi_{k}$ and $\delta\psi_{k}$ as a single degree of freedom. To verify the normalization of the solution on sub-Hubble scales, we propose the following procedure. First we compute the reduced action for the fields $\Psi_{k}$, $\Phi_{k}$ and $\delta\psi_{k}$ and we evaluate it on the solution for  $-k\tau>>1$, which corresponds to a plane wave $\sim e^{-i\omega\tau}$. Next, we identify the $\omega^2$ term in the action, after which we can already set $\omega=k$. Then, by comparing the result with that of a canonically normalized scalar field, 

\begin{equation}
\frac{1}{2}\phi'^2=Ke^{-2ik\tau}, \quad K=-\frac{k}{4} \quad\text{when}\quad \phi=\frac{e^{-ik\tau}}{\sqrt{2k}}\, ,
\end{equation}
we may be able to identify the constant $A_0$. 

First of all we need to determine the solution on sub-Hubble scales $k>>aH$.  In that case, $\delta\chi_{k}$ reads \req{eq:chiksol2}
\begin{equation}
\delta\chi_k=\frac{e^{-ik\tau}}{\sqrt{2k}}\, .
\end{equation}
On the other hand, the values of $\tilde\Psi_k$ and $\varphi_k$ can be obtained from the rest of the equations, and they read, to lowest order in $\epsilon$ and first order in  $\zeta$, 

\begin{align}
\tilde\Psi_k=&-\sqrt{\alpha \ell^2} A_0 \epsilon  \left(6i-\frac{\zeta k}{aH}\right)\frac{e^{-ik\tau}}{4 k^{3/2}}\, ,\\
\varphi_k=&-\frac{A_0}{2 aH (\alpha \ell^2)^{3/2} \epsilon\sqrt{k}  }e^{-ik\tau}\, .
\end{align}
Note that the amplitude of these fields is determined by the one of $\delta\psi_k$ \req{dpsired}, indicating that these are not independent degrees of freedom. 
Inserting this in the reduced action, we identify the following kinetic term $K$ (again, at lowest order in $\epsilon$):
\begin{equation}
K=-\frac{3A_0^2\alpha^2 \ell^4 k}{16\pi G}\, .
\end{equation}
Thus, demanding that this term is canonically normalized, $K=-k/4$, yields the normalization constant,

\begin{equation}\label{eq:A0}
A_0=\frac{1}{\sqrt{6}\alpha \ell^2 M_{\rm P}}\, .
\end{equation}
which has no corrections with respect to the $R^2$ case. 

We are now in position to compute the comoving curvature perturbation \req{comcurv} and its power spectrum on super-Hubble scales. First, note that
\begin{equation}
aH\frac{\delta\psi_k}{\psi'}\approx-\frac{\delta\psi_k}{2H^2\epsilon}\approx-\frac{3\alpha \ell^2}{2}\left(1+\zeta/9\right)\delta\psi_k
\end{equation}
Thus, on account of \req{eq:Psiksol} we have $|\Psi_k|<<|aH\frac{\delta\psi_k}{\psi'}|$, and therefore

\begin{equation}
|\mathcal{R}_{k}|\approx \frac{3\alpha}{2}\left(1+\zeta/9\right)|\delta\psi_k|\, .
\end{equation}
Then, using eqs \req{dpsired}, \req{chisupH} and \req{eq:A0} we get
\begin{equation}
|\mathcal{R}_{k}|\approx \frac{\sqrt{3}H\left(1+\zeta/9\right)}{4\sqrt{k^3}M_{\rm P}}\sqrt{1+2\alpha \ell^2 \psi}\left(\frac{k}{aH}\right)^{3/2-\nu_s}\, .
\end{equation}
Therefore, the power spectrum reads
\begin{equation}
\mathcal{P}_{S}(k)=\frac{k^3}{2\pi^2}|\mathcal{R}_{k}|^2=A_{S}^2\left(\frac{k}{aH}\right)^{n_{s}-1}\, .
\end{equation}
where the  amplitude $A_S$ and spectral index $n_{s}$ read 
\begin{align}
 A_{S}^2= & \frac{\left(1+\zeta/9\right)}{4\epsilon M_{\rm P}^2}\left(\frac{H}{2\pi}\right)^2\, ,\\
 n_{s}-1= &-4\epsilon\left(1+\frac{2\zeta}{9}\right)\, .
\end{align}

Expressed as a function of the number of e-folds we have
\begin{equation}\label{nsfinal}
n_s=1-\frac{2}{N}-\frac{32N\lambda_{3}}{27\alpha^2}\, .
\end{equation}
Again, this reproduces the result for Starobinsky inflation when $\lambda_{3}=0$. Finally, the tensor-to-scalar ratio reads
\begin{equation}\label{rfinal}
r=\frac{A_{T}^2}{A_{S}^2}=48\epsilon^2\approx \frac{12}{N^2}-\frac{64\lambda_{3}}{9\alpha^2}\, .
\end{equation}

We have again repeated this entire calculation for the quartic density \req{eq:quartic} but we refer to  Appendix~\ref{app:scalar} for details. The spectral index and tensor-to-scalar ratio including this quartic density are given by
\begin{align}
n_s &\approx 1-\frac{2}{N}-\frac{32 \lambda_3 N}{27\alpha^2}+\frac{4 \lambda_4 N^2}{3\alpha^3} \, , \label{nsfinalquartic}\\
r &\approx \frac{12}{N^2}-\frac{64\lambda_3}{9\alpha^2}+\frac{16\lambda_{4} N}{3\alpha^3} \label{rfinalquartic}	\, .
\end{align}

Thus we see that the so-called ``consistency relation'' gets modified,
\begin{equation}
\frac{r}{n_{T}}=-8+\frac{32 N^2\lambda_{3}}{3\alpha^2} - \frac{128 N^3 \lambda_4}{9 \alpha^3}\, .
\end{equation}
This shows that our model is a non-trivial modification of GR.  Note also that, even though the quartic theory \req{eq:quartic} contains three different parameters, only one of them contributes to the observables we have considered.

\subsection{Observational predictions}

\begin{figure*}[htpb]
	\begin{center}
		\includegraphics[width=0.49\textwidth]{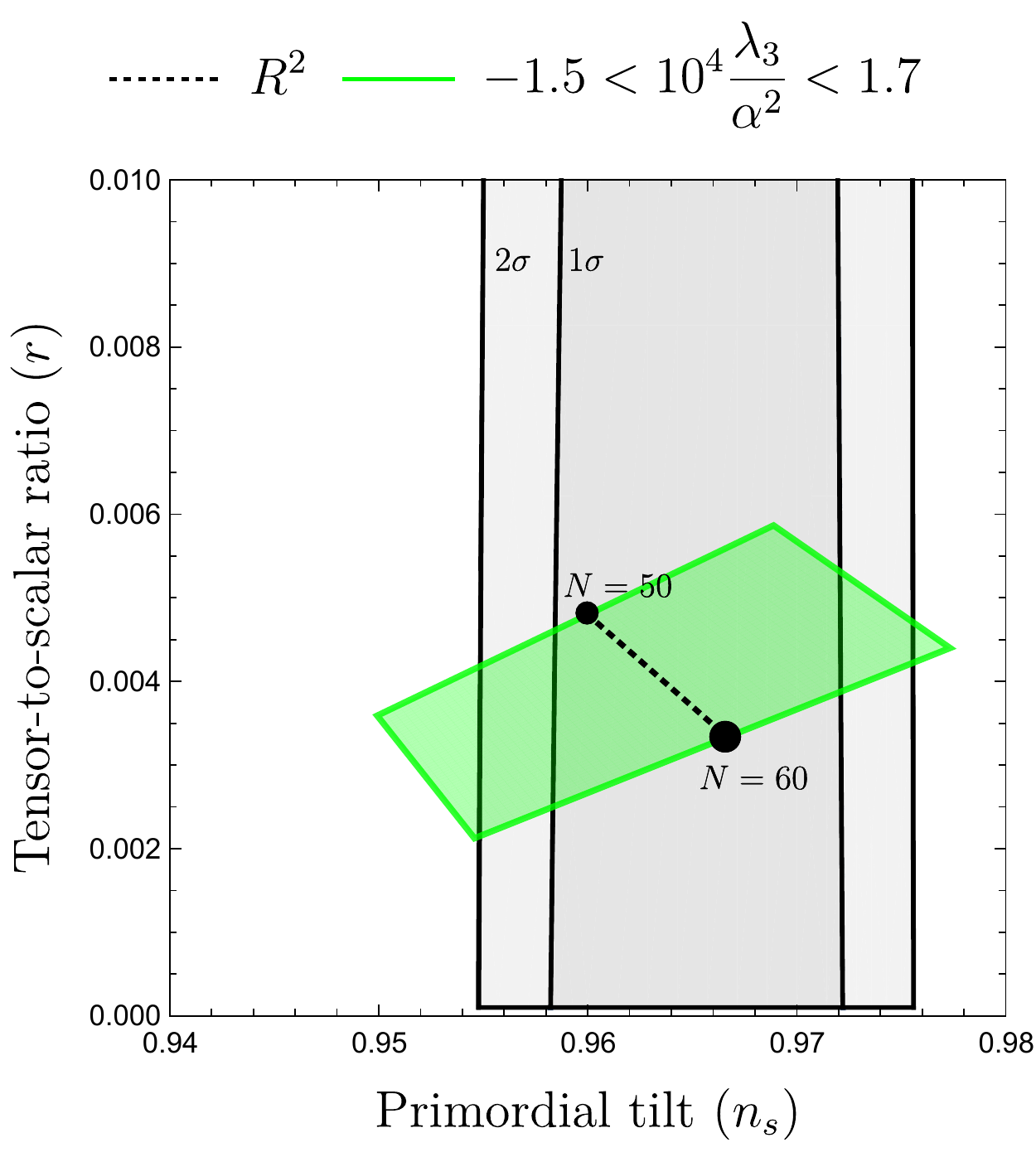}
		\includegraphics[width=0.49\textwidth]{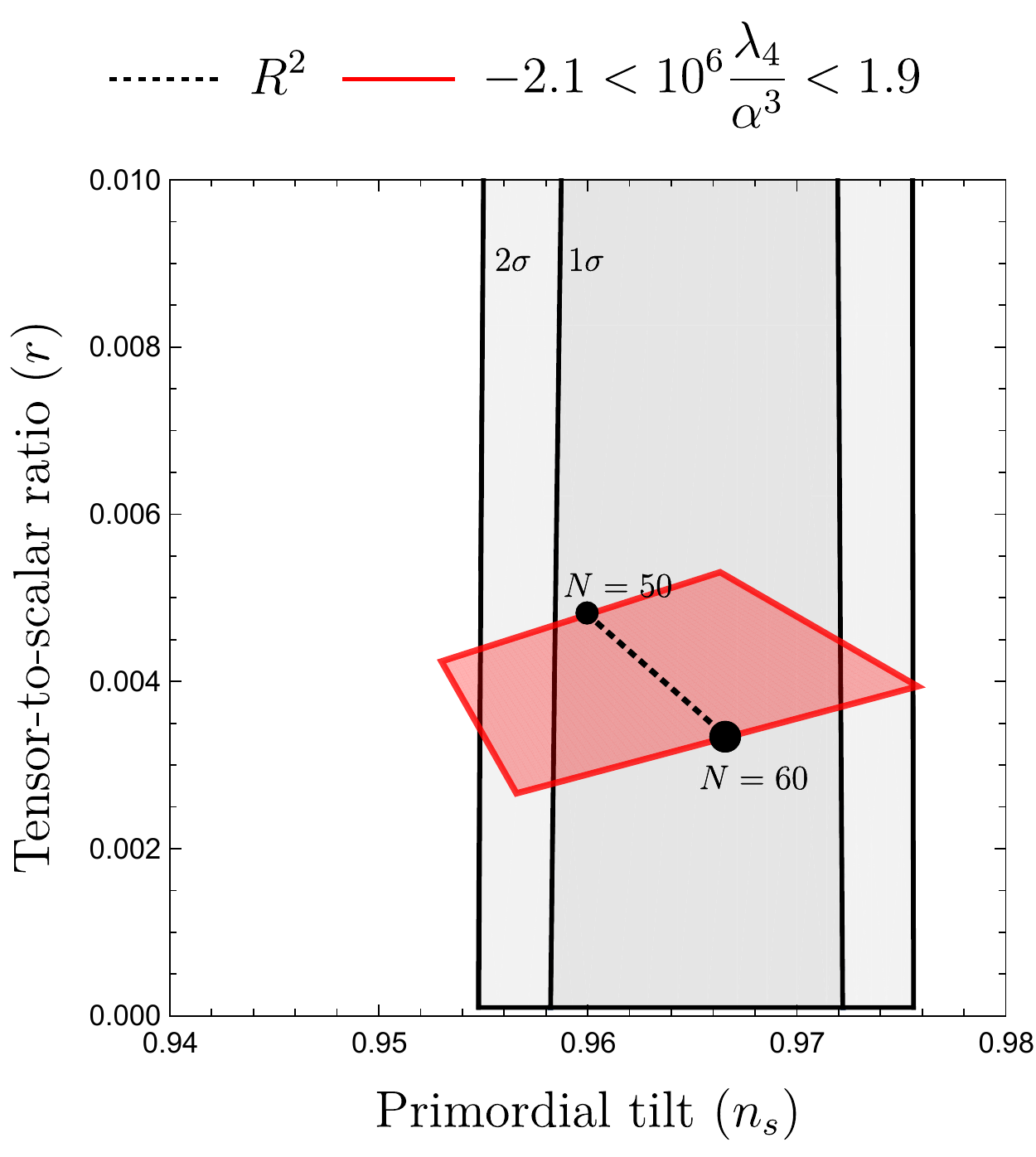}
		\caption{Predictions for the scalar spectral tilt $n_s$ and scalar-to-tensor ratio $r$ in the cubic theory \req{eq:cECG} (left) and quartic theory \req{eq:quartic} (right), given respectively in Eq.~\req{rfinalquartic} and Eq.~\req{nsfinalquartic}, for between 50 and 60 observable e-folds. Indicated are the $R^2$ ($\lambda_3=0$, $\lambda_4=0$) results (black, dotted) and the regions allowed by holographic bound \req{eq:Const2} (green for $\lambda_4 = 0$ and red for $\lambda_3 = 0$) as summarized in Table \ref{tab1}. As a reference, we also show the marginalized joint 68\% and 95\% confidence levels for these parameters as obtained by {\it{Planck}} 2018 (TT, TE, EE+lowE+lensing) for the scale $k=0.002\text{Mpc}^{-1}$ \cite{Akrami:2018odb} \footnote{Based on observations obtained with Planck (http://www.esa.int/Planck), an ESA science mission with instruments and contributions directly funded by ESA Member States, NASA, and Canada.}.}
		\label{fig:planckconstraints}
	\end{center}
\end{figure*}

We can now compare the predictions of our theories \req{eq:SRn} with the current constraints on inflationary theory from observations of CMB fluctuations. Although our formulae \req{nsfinalquartic} and \req{rfinalquartic} contain free parameters, we recall that these are bounded by the unitarity constraints we derived in Sec.~\ref{sec:holoconst}. In Fig.~\ref{fig:planckconstraints} we show the region in the $(n_s,r)$ plane that is allowed by these theoretical constraints in the range $50<N<60$. In these plots we consider separately the cases in which we have either a cubic correction or a quartic one.  As we can see, in both cases the theoretical bounds from holography on the range of values of the scalar tilt $n_s$ coincide almost exactly with the observational constraints. In the range in which the constraints are satisfied the corrections are smaller than $1$, and hence the formulae \req{nsfinalquartic} and \req{rfinalquartic} should apply.\footnote{Recall that these constraints were found by imposing unitarity for $N\le 60$. If inflation lasts longer than 60 e-folds then the constraints would be even stronger. This is relevant in particular in considerations of eternal inflation.} We see that the higher-curvature corrections do measurably change the combination $(n_s,r)$, while keeping their overall order of magnitude unchanged. The models we have considered, therefore, smear out as it were the `dot' (or the line, if you wish) corresponding to $R^2$ inflation in the $(n_s,r)$-plane. Higher-curvature corrections enrich the phenomenology of $R^2$ inflation.

\begin{figure}[t!]
	\begin{center}
		\includegraphics[width=0.49\textwidth]{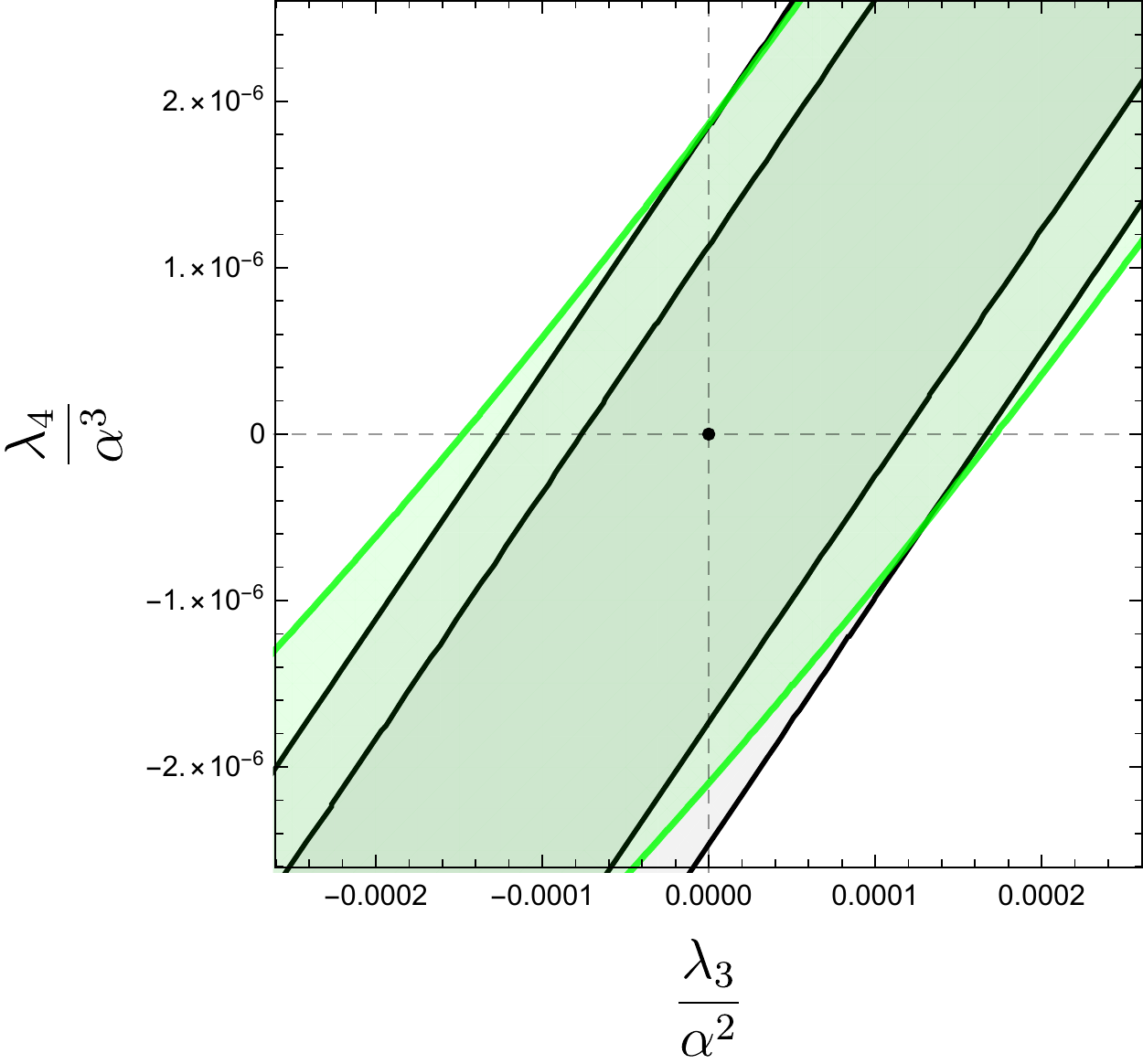} 
		\caption{Constraints on parameters $\lambda_3/\alpha^2$, associated to the cubic density \req{eq:cECG}, and $\lambda_{4}/\alpha^3$, associated to the quartic density \req{eq:quartic}, based on the holographic bound \req{eq:Const2} (green) and the marginalized joint 68\% (inner grey) and 95\% (outer grey) confidence intervals as obtained by {\it{Planck}} 2018 (TT, TE, EE+lowE+lensing) for the scale $k=0.002\text{Mpc}^{-1}$ \cite{Akrami:2018odb}, given $60$ observable e-folds of slow roll inflation. }
		\label{fig:constraintspace}
	\end{center}
\end{figure}

In theories featuring both cubic and quartic corrections the holographic constraints still allow for a narrow but infinite band in the $(\lambda_{3},\lambda_{4})$ plane --- we show this partially in  Fig.~\ref{fig:constraintspace}. One can compare this region with the bound on the parameters implied by the observations. Fig.~\ref{fig:constraintspace} shows the corresponding contour assuming $N=60$, and astonishingly we observe that it almost coincides with the holographic constraints. However, if $\lambda_{3}$ or $\lambda_{4}$ are too large, then the perturbative results \req{nsfinalquartic} and \req{rfinalquartic} cannot be applied, and for that reason in Fig.~\ref{fig:constraintspace} we restrict to values of $\lambda_{3}$ and $\lambda_{4}$ such that $|\lambda_{3}|N^2/\alpha^2< 1$, $|\lambda_{4}|N^3/\alpha^3<1$. It would be interesting to extend the results for $n_s$ and $r$ to large values of the couplings, in order to complete this comparison between the holographic constraints and experimental bounds. 

The next generation of CMB experiments will begin to probe the phenomenology of $R^2$-inflation and its higher derivative corrections. In particular, future satellite missions dedicated to the observation of the B-mode polarization are set to test the models we have presented \cite{baumann2009probing}. The (JAXA) LiteBIRD mission will, for instance, constrain $r<0.002$ at 95\% confidence level, or otherwise determine $r$ up to an error smaller than $0.001$ \cite{hazumi2019litebird}. Other proposals such as CORE \cite{Finelli:2016cyd} or PIXIE \cite{kogut2011primordial} could improve upon this, depending on the configuration. Moreover, in time this region of parameter space will be further explored by complementary ground-based experiments including CLASS \cite{essinger2014class}, POLARBEAR \cite{Adachi:2019mjv}, the South Pole Telescope \cite{Benson:2014qhw} and the next generation BICEP array \cite{Grayson:2016smb}, Simons Observatory \cite{Ade:2018sbj} and CMB-S4 \cite{Abazajian:2016yjj}. To match these exciting prospects, it would be desirable to explore the phenomenology of this class of models in more detail, e.g. by including reheating and non-Gaussianity.

\section{Discussion}\label{sec:discussion}

We have extended Starobinsky inflation with a particular class of geometric higher-curvature corrections that feature in generalized quasi-topological gravity theories. We found that a wide range of models within this class admit an extended phase of slow roll inflation as an attractor. Evaluating the semiclassical no-boundary wave function in these models, we found it predicts the universe to start out within the slow roll patch. By embedding these theories in anti-de Sitter space and using AdS/CFT, we have derived theoretical bounds on the higher-curvature terms. These constrain the leading cubic and quartic corrections to be much smaller than the coefficient of the $R^2$ term, justifying our perturbative treatment of the former. Within these bounds we have computed the leading corrections to the primordial perturbation spectrum, including the modified consistency relation $r=-8n_{T}$, and found these to be within reach of the next generation of CMB experiments. This opens up the exciting prospect to probe through future CMB observations geometric corrections to General Relativity.  

Our analysis relies heavily on the fact that the higher-curvature theories we have considered, give rise to second-order equations for cosmological backgrounds and perturbations. This renders possible a fully non-perturbative analysis of the background evolution and it means that the linearized spectrum around FLRW solutions contains no ghosts or additional modes.

However, the theories \req{eq:SRn} do not violate Lovelock's theorem. Their equations of motion are in general of higher order indeed. It is known that the reduction of the order of the equations in certain highly symmetric backgrounds usually leads to strong coupling issues. For example, for the theories we consider it was explicitly shown in \cite{Pookkillath:2020iqq,Jimenez:2020gbw} that an instability arises in anisotropic cosmological solutions.\footnote{ We note that Refs.~\cite{Pookkillath:2020iqq,Jimenez:2020gbw} only consider the cubic correction. Ref. \cite{Jimenez:2020gbw} briefly considers quartic terms but those differ from those we have studied. We have included, in principle, an arbitrary number of terms. One should not discard a priori the possibility that, when the full tower of higher-derivative terms is considered, the instabilities disappear. A similar mechanism is known to happen in infinite-derivative gravity \cite{Biswas:2011ar}.} This indicates that models of this kind must be regarded as effective theories and this has been our point of view in this paper. An EFT interpretation assumes that pathologies of the kind identified in \cite{Pookkillath:2020iqq,Jimenez:2020gbw} arise from the truncation of additional massive degrees of freedom and conjectures that when the full theory is considered, these disappear.

In any case, the most relevant results of this paper have been obtained perturbatively in the cubic and quartic couplings. These should be valid regardless one's point of view on the nature of the theory. It is worth pointing out that we have found no obvious problems, neither with the background evolution nor with the linear perturbations in our models. This suggests that the additional degrees of freedom in a UV completion of our theories may not affect the dynamics even when the couplings become large. At least, there seems to be no a priori reason they must.

Our results open up numerous avenues for further explorations of these models, both theoretically and phenomenologically. In light of the promising future observational prospects, it would be interesting to refine our predictions for observables associated with the CMB by following the evolution of the perturbations through reheating and onwards. It would also be interesting to compute the effect of the `geometric' corrections on the non-Gaussianities and to include Starobinsky's original trace anomaly terms in this analysis. Finally, especially in the context of discussions of initial conditions, it would be interesting to study the effect of the higher-curvature corrections on the regime of eternal inflation in these models. It may well be that this further constrains the range of values of the parameters.  

From a more theoretical perspective, it would be interesting to better understand the generality of our results in the context of an EFT expansion around GR. In the absence of matter, the cubic term \req{eq:cECG} is the only non-trivial correction to GR at cubic order, but matter either in the form of an $R^2$ term or as a scalar $\psi$ may change this. Nevertheless, due to the high symmetry of cosmological solutions, there are possibly not many other independent corrections at the six-derivative level. Finally it would be interesting to put the derivation of the holographic bounds on the parameters on firm footing, for instance by exploiting formulations of dS/CFT in terms of Euclidean AdS/CFT \cite{Hertog:2011ky} or by calculating these directly in de Sitter space.

\textit{Acknowledgements.}  We thank Nikolay Bobev for helpful discussions. This work is supported by the C16/16/005 grant of the KU Leuven, the COST Action GWverse CA16104, and by the FWO Grant No. G092617N. The work of PAC is supported by a postdoctoral fellowship from the FWO (12ZH121N). KF is Aspirant FWO-Vlaanderen (ZKD4846-ASP/18). 

\onecolumngrid
\appendix

\section{Details for perturbation theory}\label{app:pert}
\subsection{Equations for tensor perturbations}\label{app:tensor}
In this appendix, we extend the quadratic action for tensor perturbations \req{Sh} to include corrections from the quartic term \req{eq:quartic}
\begin{equation}
S_{h}=\frac{1}{32\pi G}\int d\tau\frac{a^2}{2}\left[\tilde{f}  h_{k}'^2-\tilde{\omega}^2 h_k^2\right]\, ,
\end{equation}
with
\begin{align}
\tilde{\omega}^2 &= \omega^2 + k^2 \left( p_1 \lambda_4 (\ell H)^6  +p_2 \frac{\lambda_3 \lambda_4 (\ell H)^8}{\alpha}+ p_3 \frac{\lambda_4^2 (\ell H)^{10}}{\alpha}-6\frac{\lambda^2_3 \lambda_4 (\ell H)^{10}}{\alpha^2}+12\frac{\lambda_3 \lambda^2_4 (\ell H)^{12}}{\alpha^2}-6\frac{\lambda_4^3 (\ell H)^{14}}{\alpha^2}\right) \\ & \quad \, + \frac{12 \lambda_4 k^4}{a^2}(H^2-\psi)\psi   \, , \notag\\
p_1 &= 394-972\epsilon+726 \epsilon^2-120 \epsilon^3-24\epsilon^4+\frac{1}{\alpha (\ell H)^2}(87-168 \epsilon+72 \epsilon^2)-\frac{6}{\alpha^2 (\ell H)^4}\, , \\
p_2&= -\frac{645}{2}+504\epsilon-168\epsilon^2+\frac{12}{\alpha (\ell H)^2}\, ,\\
p_3 &= 339-564 \epsilon+216\epsilon^2-\frac{12}{\alpha (\ell H)^2}\, , 
\end{align}
and
\begin{align}
\tilde{f} &= f +q_1 (\ell H)^{6} \lambda_4-\frac{3\lambda_3 \lambda_4 \ell^8 H^6}{2 \alpha}(H^2+8\psi) - \frac{3 \ell^{10} H^8 \lambda_4^2}{\alpha}(H^2-4\psi)+\frac{k^2}{a^2} \lambda_4 \left( 12 \ell^4 H^2(H^2-\psi) +24 \ell^4(H^2-\psi)^2 \nu \right)\, , \\
q_1 &= -38+168\epsilon-174\epsilon^2+48\epsilon^3 + \frac{3}{\alpha (\ell H)^2}(7-4\epsilon)\, . 
\end{align}
Here, $\omega$ and $f$ are respectively given in \req{eq:omega}, \req{eq:ftensor} and $\epsilon$ is still the slow roll parameter
\begin{equation}
\epsilon = 2\left(1-\frac{\psi}{H^2}\right)
\end{equation}

Using these quantities, the analysis is identical to the one presented in the main text. In particular, we first find the appropriate slow roll expressions, extending \req{Heps2} to

\begin{equation}
\ell^2 H^2 = \frac{1-\zeta/9+\xi/27}{3 \alpha \epsilon} - \frac{9+5\zeta-8\xi/3}{162 \alpha} + \cO(\epsilon, \zeta^2, \xi^2, \xi \zeta) \, ,
\end{equation}

with

\begin{equation}
\xi = \frac{\lambda_4}{\alpha^3 \epsilon^3}\, ,
\end{equation}

and keep \req{eq:aHsol} but now with

\begin{equation}
c = -2-\frac{4\zeta}{9}+\frac{2}{9} \xi \, .
\end{equation}

This leads to \req{eq:chik2} with

\begin{equation}
\tilde{k}^2 = k^2\left(1+\epsilon^2\left(\frac{3\zeta}{2}-\xi\right)\right)  \quad  \nu^2 = \frac{9}{4}+\epsilon^2\left(9+3\zeta-2\xi \right)\, .
\end{equation}

Therefore, using \req{eqn:h} and \req{eqn:PT} and \req{eqn:PTdef}

\begin{align}
A_T^2 \approx \frac{12 \epsilon(1+\zeta/9-\xi/27)}{M_P^2}\left(\frac{H}{2 \pi}\right)^2 \, , \\
n_T \approx -2\left(3+\zeta-\frac{2\xi}{3}\right)\epsilon^2\, .
\end{align}

Taking into account

\begin{equation}
\epsilon \approx \frac{1}{2 N} - \frac{4 \lambda_3}{27 \alpha^2}N+\frac{\lambda_{4}}{9 \alpha^3}N^2\, .
\end{equation}

This finally yields

\begin{equation}
n_T = -\frac{3}{2N^2}-\frac{10\lambda_{3}}{9\alpha^2} +\frac{2\lambda_4 N}{\alpha^3} \, .
\end{equation}

\subsection{Equations for scalar perturbations}\label{app:scalar}

The linearized Einstein equations that govern the scalar perturbations were summarized in Sec.~\ref{sec:scalarperturbations} as

\begin{equation}
\mathcal{E}_{\tau\tau}\, ,\quad \mathcal{E}_{\tau i}=k^i A\, ,\quad \mathcal{E}_{ij}=B k^{i}k^{j}+C g_{ij}\, .
\end{equation}

The explicit expressions for $\mathcal{E}_{\tau\tau}$, $A$, $B$, $C$, including only the additional cubic term \req{eq:cECG} in the action \req{Spsi}, are as follows

\begin{align}
\notag
\mathcal{E}_{\tau\tau}&=2 \alpha  \left(k^2+3 a^2 \left(H^2-2 \psi \right)\right) \delta \psi
_k+\left(-12 a^2 \alpha  \psi ^2+12 a^2 H^6 \lambda _3\right) \Phi
_k+\Psi _k \left(-2 k^2-4 k^2 \alpha  \psi +\lambda _3 \left(6 H^4
k^2+\frac{3 k^4 H'}{a^3}\right)\right)\\
&+6 a H \alpha  \delta \psi
_k'+\left(-6 a H+18 a H^5 \lambda _3-6 \alpha  \left(2 a H \psi +\psi
'\right)\right) \Psi _k'\, ,\\\notag
A=&2 a H \alpha  \delta \psi _k+\Phi _k \left(2 a H+\lambda _3 \left(-6 a
H^5-\frac{3 H k^2 H'}{a^2}\right)+2 \alpha  \left(2 a H \psi +\psi
'\right)\right)-2 \alpha  \delta \psi _k'\\
&+\left(2+4 \alpha  \psi
+\lambda _3 \left(-6 H^4-\frac{3 k^2 H'}{a^3}\right)\right) \Psi
_k'-\frac{3 k^2 \lambda _3 \Psi _k H''}{a^3}\, ,\\\notag
B=&2 \alpha  \delta \psi _k-\frac{9 H \lambda _3 H' \Phi _k'}{2
	a^2}+\frac{9 \lambda _3 \Psi _k' \left(a H H'-2 H''\right)}{2
	a^3}+\Phi _k \left(1+2 \alpha  \psi -\frac{3 \lambda _3 \left(2 a^3
	H^4+k^2 H'+3 a \left(\left(H'\right)^2+H H''\right)\right)}{2
	a^3}\right)\\&-\frac{9 \lambda _3 H' \Psi _k''}{2 a^3}+\Psi _k
\left(-1-2 \alpha  \psi +\frac{3 \lambda _3 \left(2 a^3 H^4+8 a^2 H^2
	H'-2 k^2 H'+3 a H H''-3 H^{(3)}\right)}{2 a^3}\right)\, ,\\\notag
C=&-2 \alpha  \delta \psi _k \left(k^2+3 a^2 \left(H^2-2 \psi \right)+2 a
H'\right)-2 a H \alpha  \delta \psi _k'+\left(2 a H+\lambda _3
\left(-6 a H^5+\frac{3 H k^2 H'}{2 a^2}\right)+2 \alpha  \left(2 a H
\psi +\psi '\right)\right) \Phi _k'\\\notag
&+\Psi _k' \left(4 a H+4 \alpha 
\left(2 a H \psi +\psi '\right)+\lambda _3 \left(-12 a H^5-24 H^3
H'-\frac{3 H k^2 H'}{2 a^2}+\frac{3 k^2 H''}{a^3}\right)\right)\\\notag
&+\Phi_k \left(6 a^2 H^2-k^2+4 a H'+\frac{3}{2} \lambda _3 \left(-12 a^2
H^6+2 H^4 \left(k^2-12 a H'\right)+\frac{k^2 H' \left(k^2+aH'\right)}{a^3}+\frac{H k^2 H''}{a^2}\right)\right. \\\notag
&\left.+\alpha  \left(-2 \psi 
\left(-6 a^2 H^2+k^2-4 a H'\right)+4 \left(a H \psi '+\psi
''\right)\right)\right)-2 \alpha  \delta \psi _k''+\left(2+4 \alpha 
\psi +\lambda _3 \left(-6 H^4+\frac{3 k^2 H'}{2 a^3}\right)\right)
\Psi _k''\\\notag
&+\Psi _k \Big(6 a^2 H^2+k^2+4 a H'+2 \alpha  \left(-6 a^2
\psi ^2+\psi  \left(6 a^2 H^2+k^2+4 a H'\right)+2 \left(a H \psi
'+\psi ''\right)\right)\\
&-\frac{3 \lambda _3}{2 a^3} \left(4 a^5 H^6+2 a^3 H^4
k^2+8 a^4 H^4 H'+8 a^2 H^2 k^2 H'+a H k^2 H''-k^2 \left(2 k^2
H'+H^{(3)}\right)\right)\Big)\, .
\end{align}

On the other hand, Eq.~\req{eq:scalarpert} reads

\begin{align}\notag
0=&\frac{3 k^2 \lambda _3 \Psi _k}{a^3} \left(-4 a^3 H^4-8 a^2 H^2 H'+k^2
H'\right)+\delta \psi _k \left(-12 a^2-6 \alpha  \left(k^2+4
a^2 \left(H^2-\psi \right)+2 a H'\right)\right)-12 a H \alpha  \delta
\psi _k'\\\notag
&+\left(-18 a H^5 \lambda _3+6 \alpha  \psi '\right) \Phi
_k'+\left(-18 \lambda _3 \left(3 a H^5+4 H^3 H'\right)+18 \alpha 
\psi '\right) \Psi _k'\\
&+\Phi _k \left(\lambda _3 \left(-72 a^2
H^6+\frac{3 k^4 H'}{a^3}+6 H^4 \left(k^2-18 a H'\right)\right)+12
\alpha  \left(2 a H \psi '+\psi ''\right)\right)-6 \alpha  \delta
\psi _k''-18 H^4 \lambda _3 \Psi _k'' \, .
\end{align}

When, in addition, we include \req{eq:quartic}, the structure is unchanged but there is an additional correction

\begin{equation}
\mathcal{E}_{\tau\tau}+\lambda_4 \mathcal{E}^{(4)}_{\tau\tau}\, ,\quad \mathcal{E}_{\tau i}=k^i (A+\lambda_4 A^{(4)})\, ,\quad \mathcal{E}_{ij}=(B+\lambda_4 B^{(4)}) k^{i}k^{j}+(C+\lambda_4 C^{(4)}) g_{ij}\, .
\end{equation}

with 

\begin{align}
\mathcal{E}^{(4)}_{\tau\tau}&= -18 a^2 H^8 \Phi_k -2k^2\frac{\Psi_k}{a^4}\left(4a^4H^6+k^2(3aH^2H'+H'{}^2)\right) -24 a H^7 \Psi'_k \, , \\
A^{(4)}&=   \frac{\Phi_k}{a^3}\left(8a^4H^7+k^2H(6aH^2H'+2H'{}^2) \right)  + \Psi'_k \left(8H^6 + \frac{k^2}{a^4}(6aH^2H'+2H'{}^2)\right) \notag\\ &+2k^2\frac{\Psi_k}{a^4}\left(2H'H''+aH(2H'{}^2+3HH'')\right)\, , \\
B^{(4)}&= 3\frac{\Phi_k'}{a^3}HH'(3aH^2+H')+3\frac{\Psi_k'}{a^4}\left(-3a^2H^3H'+4H'H''+6aH(H'{}^2+HH'')\right)  \notag \\ 
&+ \frac{\Phi_k}{a^4}\left(4H^6a^4+3aH'(H'{}^2+2HH'')+3a^2H^2(8H'{}^2+3HH'')+k^2(H'{}^2+3aH'H^2)\right) \notag \\
&+ 3\frac{\Psi_k''}{a^4}H'(3aH^2+H') + \frac{\Psi_k}{a^4} ( k^2(6aH^2-H')H'-4a^4H^6-24a^3H^4H'-3a^2H^2(2H'{}^2+3HH'') \notag \\ &+3a(2H'{}^3+6HH'H''+3H^2H''')+6(H''{}^2+H'H'''))\, , \\
C^{(4)}&= \frac{\Phi_k'}{a^3}\left(8a^4H^7-k^2HH'(3aH^2+H')\right) + \Psi_k''\left(8H^6-\frac{k^2}{a^4}H'(3aH^2+H')\right) \notag \\
&+ \Psi_k'\left( 16aH^7+48H^5H' + \frac{k^2}{a^4}(3a^2H^3H'-4H'H''-6aH(H'{}^2+HH''))\right) \notag \\
& + \Phi_k \left(8a(3aH^8+8H^6H')-\frac{k^2}{a^4}(4a^3H^6+H'{}^3+2HH'H''+aH^2(8H'{}^2+3H''))-\frac{k^4}{a^4}H'(3aH^2+H')\right) \notag \\
& + \Psi_k \left(\frac{k^2}{a^4}(ha^4H^6+24a^3H^4H'+a^2H^2(2H'{}^2+3HH'')-a(2H'{}^3+6HH'H''+3H^2H''')-2(H''{}^2+H'H'''))\right) \notag \\
&+ \Psi_k \left(2a(3aH^8+8H^6H') +\frac{k^4}{a^4}H'(-6aH^2+H')\right) \, .
\end{align}

Similarly, Eq.~\req{eq:scalarpert} becomes

\begin{equation}
6\alpha\nabla^2\psi-12\psi+\lambda_{3}\tensor{\mathcal{E}}{^{(3)\mu}_{\mu}}+\lambda_{4}\tensor{\mathcal{E}}{^{(4)\mu}_{\mu}}=0\, ,
\end{equation}

or explicitly

\begin{align}\notag
0=&\frac{3 k^2 \lambda _3 \Psi _k}{a^3} \left(-4 a^3 H^4-8 a^2 H^2 H'+k^2
H'\right)+\delta \psi _k \left(-12 a^2-6 \alpha  \left(k^2+4
a^2 \left(H^2-\psi \right)+2 a H'\right)\right)-12 a H \alpha  \delta
\psi _k'\\\notag
&+\left(-18 a H^5 \lambda _3+6 \alpha  \psi '\right) \Phi
_k'+\left(-18 \lambda _3 \left(3 a H^5+4 H^3 H'\right)+18 \alpha 
\psi '\right) \Psi _k'\\
&+\Phi _k \left(\lambda _3 \left(-72 a^2
H^6+\frac{3 k^4 H'}{a^3}+6 H^4 \left(k^2-18 a H'\right)\right)+12
\alpha  \left(2 a H \psi '+\psi ''\right)\right)-6 \alpha  \delta
\psi _k''-18 H^4 \lambda _3 \Psi _k'' \notag \\ 
&+  \frac{2k^2\lambda_4 \Psi_k}{a^4}\left(8a^4H^6+24a^3H^4H'+k^2(-3aH^2H'+2H'{}^2)\right)+\frac{2 \lambda_4 \Psi_k'}{a^4}(36a^5H^7+72a^4H^5H')+24\lambda_4 H^6 \Psi_k'' \notag \\ 
&+ \frac{\lambda_4 \Phi_k}{a^4}\left(96a^6H^8+192a^5H^6H'-8k^2a^4H^6+2k^4(-3aH^2H'-H'{}^2)\right) +24\lambda_{4} a H^7 \Phi_k'  \, .
\end{align}

The subsequent analysis of the main text can again be done mutatis mutandis. For reference, the corrected equations \req{eq:A=0}, \req{eq:B=0} and \req{eq:scalarpert2} are respectively as follows

\begin{align}
0=&\frac{4 (9-\zeta+\xi/3 )}{27 \epsilon }\left(a H  \tilde\Psi _k+ \tilde\Psi _k'\right)+ \alpha  \epsilon  \left(a H
(6+\zeta ) \delta \psi _k+(- \zeta+4\xi/9)  \delta \psi _k'\right) +\frac{4}{3} a H \alpha ^2 \epsilon  \zeta  \varphi_{k}\\
0=&\epsilon\bigg[\frac{2}{3} \alpha ^2 \zeta  \varphi_{k}+\alpha (\zeta-\frac{4\xi}{9}) \delta \psi_k+\frac{\zeta-2\xi/3}{6} \left(-2+\frac{3 k^2}{(aH)^2}\right)   \tilde\Psi _k+(\zeta-2\xi/3) \frac{\tilde\Psi _k'}{a H}+(\zeta-2\xi/3)\frac{ \tilde\Psi _k''}{2 (aH)^2}\bigg]\, , \\
0=&\delta \psi _k''+2 a H
(1+\epsilon ) \delta \psi _k'+\left(k^2-\frac{4}{3} a^2H^2 \epsilon \zeta +\frac{2}{3} a^2H^2 \epsilon \xi  \right) \delta \psi _k+\frac{  \tilde\Psi _k''}{9
	\alpha }\left((3-\epsilon ) \zeta-(4/3-2/3 \epsilon)\xi \right) \notag \\ &+\left(4 a H (6-\epsilon )+\frac{a H (84-248 \epsilon ) \zeta }{9}-\frac{a H (40-428/3 \epsilon ) \xi }{9}\right) \frac{\tilde\Psi _k'}{9 \alpha } +\left(a^2 H^2 (4-2 \epsilon )\right)\frac{\tilde\Psi _k}{\alpha} \notag \\ &+ \zeta\left(\frac{a^2 H^2(24-98 \epsilon )}{27}+\frac{k^4 \epsilon }{9 a^2 H^2  }-\frac{k^2 (-3+13 \epsilon)}{27 }\right)\frac{\tilde\Psi _k}{\alpha}- \xi\left(\frac{a^2 H^2(12-54 \epsilon )}{27}+\frac{2k^4 \epsilon }{27 a^2 H^2  }-\frac{k^2 (-4+26 \epsilon)}{81}\right)\frac{\tilde\Psi _k}{\alpha} \notag\\ &+4 a^2 H^2 \alpha  \epsilon ^2 \zeta  \varphi_k+\frac{2}{3} a H \alpha  \epsilon ^2 \zeta  \varphi_k'\, .
\end{align}

Now, with

\begin{equation}
\varphi_k \approx -\frac{(3\zeta-4\xi/3)}{2\alpha \zeta} \delta\psi_k \, , \quad \tilde{\Psi}_k \approx -\frac{3}{4}(6-\zeta/3+2\xi/3) \alpha \epsilon^2 \delta\psi_k \, ,
\end{equation}

we obtain the master equation

\begin{equation}
\delta \psi _k''+2 a H
(1+\epsilon ) \delta \psi _k'+\left(k^2-\frac{4}{3} a^2H^2 \epsilon \zeta +\frac{2}{3} a^2H^2 \epsilon \xi  \right) \delta \psi _k = 0 \, .
\end{equation}

The redefinition \req{dpsired} now implies \req{eq:chik3} with 

\begin{equation}
\nu_s^2 = \frac{9}{4}+6\epsilon\left(1+\frac{2}{9}\zeta-\frac{1}{9}\xi\right) \, .
\end{equation}

and, with $A_0$ not corrected with respect to $R^2$, we find

\begin{equation}
|\R_k| \approx \frac{\sqrt{3} H (1+\frac{\zeta}{9}-\frac{\xi}{27})}{4 \sqrt{k^3} M_P}  \sqrt{1+2\alpha \psi}\left(\frac{k}{aH}\right)^{3/2-\nu_s} \, .
\end{equation}

Therefore, the corrected amplitude $A_S$ and spectral index $n_s$ are given by

\begin{align}
A_S^2 &= \frac{1+\zeta/9-\xi/27}{4 \epsilon M_P^2}\left(\frac{H^2}{2 \pi}\right)^2 \, ,\\
n_s &=1-4\epsilon\left(1+\frac{2}{9}\zeta-\frac{1}{9}\xi\right) \approx 1-\frac{2}{N}-\frac{32 \lambda_3 N}{27\alpha^2}+\frac{4 \lambda_4 N^2}{3\alpha^3} \, .
\end{align}

Finally, the tensor-to-scalar ratio as a function of the number of e-folds reads

\begin{equation}
r = 48\epsilon^2 \approx \frac{12}{N^2}-\frac{64\lambda_3}{9\alpha^2}+\frac{16\lambda_{4} N}{3\alpha^3} 	\, .
\end{equation}

\bibliographystyle{apsrev4-1} 
\vspace{1cm}
\bibliography{Gravities} 

\end{document}